\documentclass[11pt,a4paper]{article}

\usepackage[margin=1.2in]{geometry}
\usepackage{amssymb}
\usepackage{amsmath}
\usepackage{ntheorem}
\usepackage{color}
\usepackage{graphics}
\usepackage{graphicx}
\usepackage{algorithm,algorithmic}
\usepackage{multirow}
\usepackage{algorithm,algorithmic}
\usepackage{savesym}
\savesymbol{AND}
\usepackage{epstopdf}
\usepackage{subcaption}
\usepackage{float}
\usepackage{textcomp}
\usepackage{siunitx}
\usepackage{mathtools}
\usepackage{authblk}

\DeclareMathOperator{\spn}{span}

\usepackage{tikz}
\usetikzlibrary{arrows.meta}
\usetikzlibrary{shapes.misc}

\def \O{\mathcal{O}}

\newcommand{\diagentry}[1]{\mathmakebox[1.8em]{#1}}
\newcommand{\xddots}{%
  \raise 4pt \hbox {.}
  \mkern 6mu
  \raise 1pt \hbox {.}
  \mkern 6mu
  \raise -2pt \hbox {.}
}

\newtheorem{theorem}{Theorem}
\newtheorem{remark}{Remark}
\newtheorem{lemma}{Lemma}

\newtheorem{definition}{Definition}
\newtheorem{proposition}{Proposition}

\providecommand{\keywords}[1]
{
  \small	
  \textbf{\textit{Keywords---}} #1
}

\title{Computation of invariant sets via immersion for discrete-time nonlinear systems}

\author[1]{Zheming Wang \thanks{zheming.wang@uclouvain.be}}
\author[1]{Rapha\"el M. Jungers \thanks{raphael.jungers@uclouvain.be. Rapha\"el M. Jungers is a  FNRS honorary Research Associate. This project has received funding from the European Research Council (ERC) under the European Union's Horizon 2020 research and innovation programme under grant agreement No 864017 - L2C. Rapha\"el M. Jungers is also supported by the Walloon Region and the Innoviris Foundation.}}
\author[2]{Chong-Jin Ong \thanks{mpeongcj@nus.edu.sg}}

\affil[1]{The ICTEAM Institute, UCLouvain, Louvain-la-Neuve,1348, Belgium}
\affil[2]{The Department of Mechanical Engineering, National University of Singapore, 117576, Singapore}

\date{}

\begin{document}
\maketitle

\begin{abstract}
In this paper, we propose an approach for computing invariant sets of discrete-time nonlinear systems by lifting the nonlinear dynamics into a higher dimensional linear model.  In particular, we focus on the \emph{maximal admissible invariant set} contained in some given constraint set. For special types of nonlinear systems, which can be exactly immersed into higher dimensional linear systems with state transformations, invariant sets of the original nonlinear system can be characterized using the higher dimensional linear representation. For general nonlinear systems without the immersibility property, \emph{approximate immersions} are defined in a local region within some tolerance and linear approximations are computed by leveraging the fixed-point iteration technique for invariant sets. Given the bound on the mismatch between the linear approximation and the original system, we provide an invariant inner approximation of the \emph{maximal admissible invariant set} by a tightening procedure. 
\end{abstract}


%%%%%%%%%%%%%%%%%%%%%%%%%%%%%%%%%%%%%%%%%%%%%%%%%%%%%%%%%%
\section{Introduction}
Set invariance theory is an important tool for system analysis and controller design of constrained dynamical systems, see for instance \cite{ART:Bla99,BOO:BM08} and the references therein. In particular, it is widely used in Model Predictive Control (MPC) \cite{ART:MRRS00} for systems with hard state and input constraints. In view of this, computing invariant sets becomes an active area of research and the literature is large with many different approaches developed for handling different types of systems and constraints. Since the concepts of infinite-time reachability and recursive set propagation were first introduced in \cite{ART:B72}, the fixed-point iteration technique becomes a popular framework for computing invariant sets. The early literature has been devoted to tractable fixed-point algorithms for linear systems with polyhedral constraints, see, e.g., \cite{ART:GT91,ART:Bla99} and the references therein. In the presence of bounded disturbances in linear systems, robust invariant sets are defined and corresponding fixed-point algorithms have been developed, see, e.g., \cite{ART:KG98,ART:RKKM05,ART:OG06,ART:P16}. Recently, the authors in  \cite{INP:WJO19,ART:WJO21}  have proposed a fixed-point algorithm for linear systems subject to a class of non-convex constraints. See also \cite{ART:BLAC05,ART:HO08, ART:ACFC09,ART:FAC10,ART:SG12} for fixed-point algorithms computing invariant sets of nonlinear systems. The fixed-point iteration technique by its nature is only applicable to discrete-time systems. For continuous-time systems, there also exist algorithms for computing invariant sets, see, e.g., \cite{ART:HK14,ART:KHJ14}. However, obtaining an exact invariant set remains a challenging problem for general nonlinear systems. The aforementioned algorithms for nonlinear systems focus on inner or outer approximations of invariant sets except for special systems. However, these approximations are not necessarily invariant. In this paper, we attempt to characterize invariant sets by using a lifted linear model of the nonlinear system.

Computing linear equivalents or approximations of nonlinear systems is one of the most well-known research topics in systems and control. Classic linearization methods like Jacobian linearization and feedback linearization can be found in \cite{BOO:K02}. A more advanced linearization method is the state immersion method, which immerses a nonlinear system into a linear system in a higher dimension, see, e.g., \cite{ART:MN83,ART:LM88, ART:MT09}. 
Although the immersion method is equivalent to the feedback linearization method in the special case where the immersion is a diffeomorphism, they are in general different as an immersion does not necessarily preserve the dimension of the system. Successful applications of state immersion can be found in observer design and output regulation of nonlinear system \cite{ART:KI83,ART:LM86,ART:BPIK97}. Recently, a new immersion technique has been proposed in \cite{INP:JT19} for continuous-time systems by the use of polyflows. While a nilpotency property is required for the exact immersion or linearization, the approximation by polyflows often outperforms the Taylor approximation in practice. Inspired by the polyflows approximation, we have developed a similar immersion method \cite{ART:WJ20} for discrete-time systems.  In this paper, we use such a method to obtain a high-dimensional linear model for the characterization of invariant sets of discrete-time nonlinear systems.  Let us add that, while the goal is quite different, the immersion method in \cite{ART:WJ20} bears some similarities with classic identification techniques \cite{BOO:L87}. For example, it is similar to multivariate autoregressive modeling \cite{ART:HPF03} except that we provide the connection between the linear model and the immersibility property of the system.

Operator-theoretic approaches like the Carleman linearization \cite{BOO:KS91} and the Koopman approach (see, e.g., \cite{ART:WKR15} and the references therein) are also promising frameworks to provide an (infinite-dimensional) linear representation of nonlinear systems. For numerical analysis, the infinite-dimensional linear operator is often truncated into finite-dimensional approximations, which then can be used for (global) system analysis and prediction. For instance, in \cite{ART:MM16}, Koopman eigenfunctions are computed to characterize invariant sets of the system. While these characterizations are quite useful for stability analysis, they are in general not optimal with respect to the constraint set. In particular, the \emph{maximal admissible invariant set}, which is the maximal invariant set contained in a given constraint set, can not be easily computed from Koopman eigenfunctions. In this paper, we explicitly take the constraint set into account in two ways: First, the fixed-point iteration technique for invariant sets is used in the computation of the lifted linear model. Second, a fixed-point algorithm in the lifted space is designed to ensure set invariance inside the constraint set with a tightening procedure that circumvents mismatch error of the immersion.

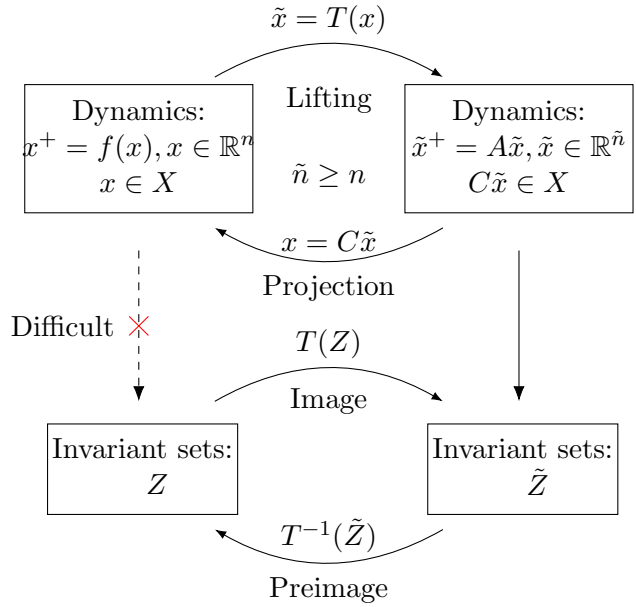
\begin{figure}[h]
\centering
\tikzset{cross/.style={cross out, draw=red, 
         minimum size=2*(#1-\pgflinewidth), 
         inner sep=0pt, outer sep=0pt}}
\begin{tikzpicture}
\draw[draw=black] (11,5) rectangle ++(3,1.7) node[pos=.5] {\begin{tabular}{c} Dynamics: \\ 
 $x^+=f(x), x\in \mathbb{R}^n$\\
 $x\in X$
\end{tabular}};
\draw[draw=black] (16,5) rectangle ++(3,1.7) node[pos=.5] {\begin{tabular}{c} Dynamics: \\ 
 $\tilde{x}^+=A\tilde{x}, \tilde{x}\in \mathbb{R}^{\tilde{n}}$\\
 $C\tilde{x} \in X$
\end{tabular}};
\draw [-{Latex[length=1.5mm]}] (13.5,6.8) to [bend left=30] node [above, sloped]  {$\tilde{x}=T(x)$} (16.5,6.8);
\draw [{Latex[length=1.5mm]}-] (13.5,4.8) to [bend right=30] node [above, sloped]  {$x=C\tilde{x}$} (16.5,4.8);
\node at (15,6.5) {Lifting};
\node at (15,4) {Projection};
\draw[draw=black] (11.3,1) rectangle ++(2.5,1.2) node[pos=.5] {\begin{tabular}{c} Invariant sets: \\ 
 \quad $Z$
\end{tabular}};
\draw[draw=black] (16.3,1) rectangle ++(2.5,1.2) node[pos=.5] {\begin{tabular}{c} Invariant sets: \\ 
 \quad $\tilde{Z}$
\end{tabular}};
\draw [-{Latex[length=1.5mm]}] (13.5,2.5) to [bend left=30] node [above, sloped]  {$T(Z)$} (16.5,2.5);
\draw [{Latex[length=1.5mm]}-] (13.5,0.8) to [bend right=30] node [above, sloped]  {$T^{-1}(\tilde{Z})$} (16.5,0.8);
\node at (15,2.5) {Image};
\node at (15,0) {Preimage};
\node at (15,5.5) {$\tilde{n}\ge n$};
\draw[dashed,-{Latex[length=2mm]}] (12.5,4.5) -- (12.5,2.5); 
\draw  (12.5,3.5) node[cross=4,rotate=0]{ };
\node at (11.5,3.5) {Difficult};
\draw[-{Latex[length=2mm]}] (17.5,4.5) -- (17.5,2.5);
%\node at (18.5,3.5) {Easy};
\end{tikzpicture}
\caption{Invariant set computation via immersion: a linear representation of the nonlinear system enables tractable characterizations of invariant sets.}\label{fig:basicimm}
\end{figure}

Inspired from many different linearization techniques in different fields, including the Koopman approach \cite{ART:WKR15}, Carleman linearization \cite{BOO:KS91}, polyflow approximation \cite{INP:JT19} and embedding theorems like Taken's theorem \cite{INC:T81}, we derive an immersion-based approach for invariant set computation of nonlinear systems. The basic idea of our approach is illustrated in Figure \ref{fig:basicimm}. When an exact immersion is available, invariant sets of the original nonlinear system can be computed using the lifted linear model. This paper focuses on the \emph{maximal admissible invariant set} contained in some given constraint set. Our contribution is threefold. First, we formally introduce the concept of \emph{approximate immersions} for general nonlinear systems and use it for characterizing invariant sets. While this concept seems quite natural, it has not been formally mentioned in the literature. Most importantly, it allows us to construct invariant inner approximations of the \emph{maximal admissible invariant set}, provided that a bound on the mismatch error between the original system and the lifted linear system is computed. Second, we leverage the fixed-point iteration technique for invariant sets to compute \emph{approximate immersions} and derive convergence properties. Third, we show that, for special classes of nonlinear systems, this approach produces the exact \emph{maximal admissible invariant set}.

The rest of the paper is organized as follows. This section ends with the notation,  followed by the next section on the review of preliminary results on invariant sets. In Section \ref{sec:main}, we will discuss the immersibility property and the immersion method using the the fixed-point iteration technique for invariant sets. Section \ref{sec:invimm} presents the proposed immersion-based method for computing the \emph{maximal invariant set} of nonlinear systems. Some computational aspects of the proposed method will be discussed in Section \ref{sec:com}. Numerical examples are provided Section \ref{sec:num}. The last section concludes the work.

A preliminary version of this paper appears as a conference paper in \cite{ART:WJO20}, which relies on the assumption that the system is asymptotically stable at the origin. This assumption is now relaxed in this paper, which leads to significant changes in the proofs of the main results. In addition, we provide a special family of polynomial systems in which an exact immersion can be obtained.

\textbf{Notation}. The non-negative real number set and the non-negative integer set are indicated by $\mathbb{R}^+$ and $\mathbb{Z}^+$ respectively.  $I_n$ is the $n\times n$ identity matrix and $\pmb{0}_{n\times m}$ is the $n\times m$ matrix of all zeros (subscript omitted when the dimension is clear). $\mathbb{B}_n$ is the unit closed ball in $\mathbb{R}^n$. $\|x\|_{p}$ denotes the $\ell_{p}$-norm of $x$($\|x\|=\|x\|_2$ by default) and $\|x\|_F$ is the Frobenius norm. Given a set $S$ and a vector $x$, $\pmb{1}_S$ denotes the indicator function of $S$ and $\textrm{dist}(x,S)$ denotes the distance from $x$ to $S$, defined by $\textrm{dist}(x,S) = \inf_{y\in S} \|x-y\|$. For two set $X$ and $Y$, $X\ominus Y$ denotes the Minkowski difference. Given a map $T$, let $T(X)$ denote $\{T(x):x\in X\}$ and $T^{-1}(Y)$ denote the preimage of the set $Y$ under the map $T$, i.e., $T^{-1}(Y)\coloneqq\{x:T(x)\in Y\}$ ($T$ is not necessarily invertible). A function $\alpha: \mathbb{R}^+ \rightarrow \mathbb{R}^+$ is of class $\mathcal{K}$ if it is continuous and strictly increasing with $\alpha(0)=0$. Given a list of column vectors $\{x_i\}_{i=1}^N$, $(x_1,x_2,\cdots, x_N)$ denotes the stacked vector $[
x_1^\top ~  x_2^\top ~ \cdots ~ x_N^\top
]^\top$. 

%
% For any two matrices $A,B$, $A\otimes B$ denotes the Kronecker product and $A^{[k]}=A  \underbrace{\otimes \cdots \otimes}_{k ~ times} A$ for $k\in \mathbb{Z}^+$. 

\section{Preliminaries}\label{sec:pre}
We consider discrete-time dynamical systems of the form
\begin{align}
x(t+1) &= f(x(t)), \quad t\in \mathbb{Z}^+ \label{eqn:fx}.
\end{align}
where $x(t)\in \mathbb{R}^n$ is the state vector and $f: \mathbb{R}^n \rightarrow \mathbb{R}^n$ is a continuous function. The system is subject to state constraints:
\begin{align}
x(t) &\in X \subseteq \mathbb{R}^n, \quad t\in \mathbb{Z}^+. \label{eqn:X}
\end{align}
The goal of this paper is to compute an invariant set of System (\ref{eqn:fx}) inside $X$. The formal definition of invariant sets is given below, see, e.g., \cite{ART:Bla99,BOO:BM08}.
\begin{definition}\label{def:cainv} A nonempty set $Z\subseteq \mathbb{R}^n$ is a \emph{positively invariant set} for System (\ref{eqn:fx}) if $x\in Z$ implies $f(x)\in Z$.
\end{definition}

Invariant sets throughout the paper are all positively invariant sets. Computing an invariant set can be a difficult even for linear systems, depending on the constraint set $X$, see,e.g., \cite{ART:WJO21}. For nonlinear systems, the computation is more difficult and complicated. For instance, let us consider the computation of the \emph{maximal admissible invariant set} \cite{ART:GT91,ART:KG98}, which is defined below.

\begin{definition}\label{def:cainvmax}
A nonempty set $S$ is the \emph{maximal admissible invariant set} for System (\ref{eqn:fx}) if $S$ is an invariant set and contains all the invariant sets inside $X$. 
\end{definition}

The \emph{maximal admissible invariant set} can be determined by the following fixed-point iteration (see, e.g., \cite{ART:B72,ART:GT91})
\begin{align}
O_0 \coloneqq X, 
O_{k+1} \coloneqq O_k \bigcap \{x: f(x)\in O_k\}, k \in \mathbb{Z}^+ . \label{eqn:Ok}
\end{align}
Thus, the \emph{maximal admissible invariant set} is given by
\begin{align}\label{eqn:Oinf}
O_{\infty} \coloneqq \lim_{k\rightarrow \infty} O_k.
\end{align}
While the set $O_{\infty}$ can be computed efficiently for linear systems with linear constraints by solving linear optimization problems, see, e.g., \cite{ART:Bla99,ART:GT91}, in general, it is challenging because one has to solve non-convex optimization problems. In this paper, we attempt to tackle the issue of nonlinearity via state immersion. More precisely, we propose to use a lifted linear model to compute invariant sets.

The following assumptions are made. (\textbf{A1}) The function $f(x)$ is Lipschitz continuous in $X$ with a Lipschitz constant $L_f$. (\textbf{A2}) The set $X$ is compact and contains an invariant set with a non-empty interior. (\textbf{A3}) There exist a set $\mathcal{A}$ and a class $\mathcal{K}\mathcal{L}$ function $\beta$ such that $\mathcal{A}+\epsilon \mathbb{B}_n \subset X$ for some $\epsilon>0$ and $
|f^t(x)|_{\mathcal{A}} \le \beta(|x|_{\mathcal{A}},t), \forall t\in \mathbb{Z}^+, \forall x\in X, 
$
where $
f^{t}(x) = f(f^{t-1}(x))$ with $f^0(x) = x
$. The last assumption means that System (\ref{eqn:fx}) is uniformly asymptotically stable with respect to $\mathcal{A}$ in $X$. We refer the reader to \cite{ART:A02, ART:K14} for definitions of uniform asymptotic stability and class $\mathcal{K}\mathcal{L}$ functions.

With the assumptions above, the convergence properties of the fixed-point iteration in (\ref{eqn:Ok}) are stated in the following proposition.

\begin{proposition}\label{prop:Oinfnl}
Consider System (\ref{eqn:fx}) with the constraint set $X$ as defined in (\ref{eqn:X}), let $O_k$ be defined in (\ref{eqn:Ok}) for any $k\in \mathbb{Z}^+$. Suppose (\textbf{A1}) $\&$ (\textbf{A2}) hold, then the following properties hold. (i) $O_{\infty}$ contains a non-empty interior.(ii) For any $k\in \mathbb{Z}^+$, $O_{k}$ is compact. (iii) If (\textbf{A3}) also holds, there exists a finite $k^*$ such that
$
O_{k}=O_{k^*}
$
for all $k \ge k^*$ and $O_{\infty} = O_{k^*}$. 
\end{proposition}
Proof: The proof is adapted from Proposition 3 in \cite{ART:WJO21}. (i)  This is a direct consequence of  (\textbf{A1}) and (\textbf{A2}). (ii) From (\ref{eqn:Ok}), for all $k\in \mathbb{Z}^+$, $O_k$ can be written as $O_k = \{x: f^\ell(x) \in X, \ell =0,1,\cdots, k\}$. As $f(x)$ is continuous in $X$ and $X$ is compact, $O_k$ is also compact for any $k\in \mathbb{Z}^+$. (iii)  (\textbf{A3}) implies that there exists $k'$ such that $f^{k'}(x) \in X$ for all $x\in X$, as $\mathcal{A}+\epsilon \mathbb{B}_n$ is contained in $X$ for some $\epsilon>0$.  Hence, $O_{k'} = \{x: f^\ell(x) \in X, \ell =0,1,\cdots, k'\} =  \{x: f^\ell(x) \in X, \ell =0,1,\cdots, k'-1\} = O_{k'-1}$. Following the same arguments in Proposition 3 in \cite{ART:WJO21}, Property (ii) can be proved with $k^* = k'-1$. $\Box$

%(\textbf{A4}) The function $f(x)$ is Lipschitz continuous in $X$ with a Lipschitz constant $L$: 
%\begin{align}
%\|f(x)-f(y)\|\le L\|x-y\|, \quad \forall x,y \in X.
%\end{align}
%(\textbf{A5}) The disturbance set $W$ is compact and contains the origin its interior and a bound $D$ is available: $W \subseteq D \mathbb{B}_p$.

Property (iii) of Proposition \ref{prop:Oinfnl} is called the finite determinability property, which means that $O_{\infty}$ can be computed in a finite number of steps. This property is first introduced in \cite{ART:B72} and is formalized later in \cite{ART:GT91}. More discussions on this property can be also found in \cite{ART:HO08} for certain nonlinear systems.

%\begin{remark}
%The assumption \textbf{A3} guarantees the existence of a nonempty interior of the \emph{maximal admissible invariant set}. 
%\end{remark}

\section{Immersion and approximate immersion}\label{sec:main}
This section discusses the computation of linear equivalents and approximate linear equivalents of nonlinear systems via immersion.

%In this section, we characterize the maximal invariant set $O_{\infty}$ of the system (\ref{eqn:fx}) by an equivalent or approximate linear model. We will first discuss state immersion and then present some set invariance properties under state immersion. For general nonlinear systems, we can only achieve approximate lifted linear models because an exact immersion to a linear system may not exist. For this reason, we need to take the system mismatch and robustness into consideration in the proposed characterization.

\subsection{State immersion}
First, we recall the definition of immersibility of nonlinear systems, see, e.g., \cite{ART:MN83,ART:LM88}.
\begin{definition}\label{def:imm}
System (\ref{eqn:fx}) is \emph{immersible} into a linear system in the form of 
\begin{align}
\xi(t+1) = A_{\xi}\xi(t), ~y(t)  = C_{\xi} \xi(t), ~ t \in \mathbb{Z}^+ \label{eqn:xiy2},
\end{align}
where $\xi\in \mathbb{R}^{n_{\xi}}$, $y(t) \in \mathbb{R}^n$, $A_{\xi} \in \mathbb{R}^{n_{\xi}\times n_{\xi}}$ and $C_{\xi}\in \mathbb{R}^{n\times n_{\xi}}$, if there exists a map $T: \mathbb{R}^n \rightarrow \mathbb{R}^{n_{\xi}}$ such that
$
T(f(x)) = A_{\xi}T(x),  C_{\xi}T(x) = x, \forall x\in \mathbb{R}^n.
$
For notational simplicity, let us denote the linear system in (\ref{eqn:xiy2}) by $\Pi(A_{\xi},C_{\xi})$.
\end{definition}

This definition means that, when the original system in (\ref{eqn:fx}) is immersible to some linear system, the trajectory can be considered as a linear projection of this high-dimension linear system under some transformation map. A necessary and sufficient condition for immersibility is given in the following proposition.
\begin{proposition}\label{prop:imm}
System (\ref{eqn:fx}) is immersible into a linear system in the form of (\ref{eqn:xiy2}) if and only if there exist $M$ and a sequence of matrices $\{\gamma_{\ell}\in \mathbb{R}^{n\times n}\}_{\ell=0}^M$ such that
\begin{align}\label{eqn:fMplus1}
f^{M+1}(x) = \sum\limits_{\ell=0}^M \gamma_{\ell}f^{\ell}(x), \forall x\in \mathbb{R}^n
\end{align} 
\end{proposition}
\vspace{-3mm}
The proof can be found in \cite{ART:WJ20} and similar arguments can also be found in \cite{ART:MN83,ART:LM88}. With the integer $M$ and the matrices $\pmb{\gamma}_M: =\{\gamma_{\ell}\in \mathbb{R}^{n\times n}\}_{\ell=0}^M$ satisfying (\ref{eqn:fMplus1}), we can immediately construct a linear system $\Pi(\Gamma(\pmb{\gamma}_M),[I_n~\pmb{0}_{n\times Mn}])$, where 
\begin{align}\label{eqn:GammagammaM}
\Gamma(\pmb{\gamma}_M)\coloneqq\left(\begin{array}{ccccc}
\pmb{0} & I_n & \pmb{0} & \cdots & \pmb{0}\\
\vdots & \vdots & \vdots & \vdots & \vdots\\
\pmb{0} & \pmb{0} & \cdots & \pmb{0}  & I_n\\
\gamma_0 & \gamma_1 & \cdots & \gamma_{M-1} & \gamma_{M}
\end{array}
\right).
\end{align}
The condition in (\ref{eqn:fMplus1}) implies that System (\ref{eqn:fx}) is immersible into $\Pi(\Gamma(\pmb{\gamma}_M),[I_n~\pmb{0}_{n\times Mn}])$ with the transformation map
\begin{align}\label{eqn:FMx}
 \mathcal{F}_M(x)\coloneqq \left( x, f(x), \cdots, f^M(x) \right)
\end{align}
However, there may exist redundancy in such a transformation. To remove redundancy, we will use \emph{linearly independent} transformations, defined below.

\begin{definition}
Given a nonempty set $S\subseteq \mathbb{R}^n$, a map $T: \mathbb{R}^n \rightarrow \mathbb{R}^{m}$ is called \emph{linearly independent} in $S$ if $\spn\{T(x):x\in S\} = \mathbb{R}^m$. 
\end{definition}

%From this definition, a \emph{linearly independent} map $T: \mathbb{R}^n \rightarrow \mathbb{R}^{m}$ implies $\spn\{T(x):x\in \mathbb{R}^n\} = \mathbb{R}^m$. Otherwise, there exists a vector $c\in \mathbb{R}^m$ such that $c^TT(x) = 0, \forall x\in \mathbb{R}^n$.

With a \emph{linearly independent} transformation, a tight linear model can be obtained, as stated in the following theorem. 
\begin{theorem}\label{thm:immT}
Suppose (\textbf{A1}) $\&$ (\textbf{A2}) hold and System (\ref{eqn:fx}) is immersible into a linear system in the form of (\ref{eqn:xiy2}). Let $O_{\infty}$ be defined as in (\ref{eqn:Oinf}). Then, there always exist a continuous \emph{linearly independent} map $T: \mathbb{R}^n \rightarrow \mathbb{R}^{m}$ in $O_{\infty}$ and an observable pair $(C,A)$ such that $AT(x) = T(f(x))$ and $CT(x) = x$ for all $x\in O_{\infty}$, and the trajectories of $\Pi(A,C)$ are always bounded, i.e., $\sup_{t\in \mathbb{Z}^+} \|A^t\| < \infty$. Moreover, if \textbf{A3} also holds with $\mathcal{A}=\{0\}$ and $f(0)=0$, $A$ is Schur stable. 
\end{theorem}
Proof: From Proposition \ref{prop:imm}, when System (\ref{eqn:fx}) is immersible to a linear system, there exist $M$ and matrices $\pmb{\gamma}_M\coloneqq\{\gamma_{\ell}\in \mathbb{R}^{n\times n}\}_{\ell=0}^M$ such that (\ref{eqn:fMplus1}) is satisfied. This implies that 
\begin{align}\label{eqn:fMplus1Gamma}
\mathcal{F}_M(f(x))
=\Gamma(\pmb{\gamma}_M) \mathcal{F}_M(x)
\end{align}
where $\Gamma(\pmb{\gamma}_M)$ is defined in (\ref{eqn:GammagammaM}). Suppose there are $m$ \emph{linearly independent} functions that form a basis for the span of  $\{x_1,\cdots, x_n,\cdots, f^M_1(x),\cdots, f^M_n(x)\}$ in $O_{\infty}$, let $T(x)$ be the stacked vector of these functions. As all the functions $\{x_1,\cdots, x_n,\cdots, f^M_1(x),\cdots, f^M_n(x)\}$ can be expressed as linear combinations of $T(x)$, there exists a full column rank matrix $P\in \mathbb{R}^{(M+1)n\times m}$ such that
$
\mathcal{F}_M(x) = PT(x), \mathcal{F}_M(f(x))= PT(f(x)).
$
Hence, from (\ref{eqn:fMplus1Gamma}), $T(f(x)) = P^+ \Gamma(\pmb{\gamma}_M)PT(x)$, where $P^+$ denotes the pseudo inverse of $P$. Letting $A=P^+ \Gamma(\pmb{\gamma}_M)P$ and $C = [I_n~\pmb{0}_{n\times Mn}]P$, we can get $AT(x) = T(f(x))$ and $CT(x) = x$. When $ T(x) = \mathcal{F}_M(x)$, $A=\Gamma(\pmb{\gamma}_M)$ and $C = [I_n~\pmb{0}_{n\times Mn}]$. From the definition of $\Gamma(\pmb{\gamma}_M)$, it can be immediately verified that $([I_n~\pmb{0}_{n\times Mn}],\Gamma(\pmb{\gamma}_M))$ is observable. Now, we will show that $(C,A)$ is observable for the case where $T(x)$ is not the whole $\mathcal{F}_M(x)$. As $T(x)$ is \emph{linearly independent} in $O_{\infty}$, we can choose $m$ points $\{\tilde{x}^1,\tilde{x}^2,\cdots, \tilde{x}^m\}$ inside $O_{\infty}$ such that $\spn\{T(\tilde{x}^1),T(\tilde{x}^2),\cdots, T(\tilde{x}^m)\} = \mathbb{R}^m$. Hence, for any $z\in \mathbb{R}^m$, there exist $\{\alpha_1,\alpha_2,\cdots, \alpha_m\}$ such that $z=\sum_{i=1}^m\alpha_i T(\tilde{x}^i)$. Thus, with some manipulations, it can be shown that $CA^{\ell}z = (I_n~\pmb{0}_{n\times Mn}) \Gamma(\pmb{\gamma}_M)^{\ell}Pz$ for all $\ell = 0,1,\cdots,M$, which implies that
\begin{align*}
\left( \begin{array}{c}
Cz\\
%CAz\\
\vdots\\
CA^{M}z
\end{array} \right) = \left( \begin{array}{c}
(I_n~\pmb{0}_{n\times Mn}) Pz\\
%(I_n~\pmb{0}_{n\times Mn}) \Gamma(\pmb{\gamma}_M)Pz\\
\vdots\\
(I_n~\pmb{0}_{n\times Mn}) \Gamma(\pmb{\gamma}_M)^{M}Pz\
\end{array} \right)=Pz
\end{align*}
Since $P$ is full column rank, the vector $z$ can be also uniquely determined by the output sequence $\{Cz,CAz,\cdots, CA^{M-1}z\}$. This holds for any $z\in \mathbb{R}^m$. Therefore, we conclude that $(C,A)$ is observable.  Now, we show that $A^t$ is bounded for any $t\in \mathbb{Z}^+$. Since, given any $z\in \mathbb{R}^m$, it can be written as $z=\sum_{i=1}^m\alpha_i T(\tilde{x}^i)$ for some $\{\alpha_1,\alpha_2,\cdots, \alpha_m\}$. Then, $A^tz =\sum_{i=1}^m\alpha_i A^tT(\tilde{x}^i)  = \sum_{i=1}^m\alpha_i T(f^t(\tilde{x}^i))$ for any $t\in \mathbb{Z}^+$. From the invariance of $O_{\infty}$, $f^t(\tilde{x}^i)\in O_{\infty}$ for any $i=1,2,\cdots, m$ and $t\in \mathbb{Z}^+$, which implies the boundedness of $A^tz$. Hence, $A^t$ is bounded any $t\in \mathbb{Z}^+$. When \textbf{A3} holds with $\mathcal{A}=\{0\}$, $\lim_{t\rightarrow \infty}f^t(\tilde{x}^i) = 0$ for any $i=1,2,\cdots,m$. Hence, $\lim_{t\rightarrow \infty} A^tz = 0$, implying that $A$ is asymptotically stable and thus Schur stable. $\Box$

\subsection{A special family of polynomial systems}\label{sec:special}
For certain classes of nonlinear systems, exact finite-dimensional immersions are guaranteed, see, e.g., a few classes of continuous-time nonlinear systems given in \cite{ART:LM86,ART:W83}. In this section, we exhibit another class of discrete-time polynomial systems which admit finite-dimensional immersions. Consider polynomial systems in the form of
\begin{subequations}\label{eqn:xyz}
\begin{align}
\eta(t+1)& = A_\eta \eta(t) + \varphi(z(t))\\
z(t+1)& = A_z z(t), ~~ t\in \mathbb{Z}^+
\end{align}
\end{subequations}
where $\eta \in \mathbb{R}^{n_\eta},z\in \mathbb{R}^{n_z}$, $A_\eta\in \mathbb{R}^{n_\eta\times n_\eta}$, $A_z\in \mathbb{R}^{n_z\times n_z}$,  and $\varphi: \mathbb{R}^{n_z} \rightarrow \mathbb{R}^{n_\eta}$ is a polynomial function of degree $d\in \mathbb{Z}^+$. Let us consider the the algebraic lifting in \cite{ART:BN05,ART:PJ08}. Given any $z\in \mathbb{R}^{n_z}$ and $d\in \mathbb{Z}^+$, let $z^{[d]}$ denote the $d$-lift of $z$ which consists of all possible monomials of degree $d$, indexed by all the possible exponents $\alpha$ of degree $d$
$
z^{[d]}_{\alpha} = \sqrt{\alpha !} z^{\alpha}
$
where $\alpha = (\alpha_1,\cdots,\alpha_n)$ with $\sum_{i=1}^n \alpha_i=d$ and $\alpha !$ denotes the multinomial coefficient 
$
\alpha ! \coloneqq \frac{d!}{\alpha_1 ! \cdots \alpha_n !}.
$
The $d$-lift of the matrix $A_z\in \mathbb{R}^{n_z\times n_z}$ is defined as: $A_z^{[d]}: z^{d} \rightarrow (A_z z)^{[d]}$. With a slight abuse of notation, let 
\begin{align}\label{eqn:zAd}
z^{[\pmb{d}]} \coloneqq \begin{pmatrix}
z^{[1]}\\
%z^{[2]}\\
\vdots\\
z^{[d]}\\
\end{pmatrix}, 
A_z^{[\pmb{d}]} \coloneqq \begin{pmatrix}
    \diagentry{A_z^{[1]}}\\
%    &\diagentry{A_z^{[2]}}\\
    &\diagentry{\xddots}\\
    &&\diagentry{A_z^{[d]}}\\
\end{pmatrix}
\end{align}
where $\pmb{d} \coloneqq \{1,2,\cdots, d\}$. With these definitions, $\varphi(z)$ can be expressed as
$
\varphi(z): = F_1z^{[1]}+ \cdots + F_dz^{[d]} = Fz^{[\pmb{d}]}$ 
with $F_i \in \mathbb{R}^{n_\xi\times {n_z+i-1 \choose i}}$ for $i= 1,2,\cdots, d$ and $F\coloneqq [ F_1 ~ F_2 ~ \cdots ~ F_d ]$.  The immersibility property of System (\ref{eqn:xyz}) is then stated in the following theorem. 
\begin{theorem}\label{thm:special}
Consider System (\ref{eqn:xyz}) with $\varphi: \mathbb{R}^{n_z} \rightarrow \mathbb{R}^{n_\eta}$ being a polynomial function of degree $d\in \mathbb{Z}^+$, given by $\varphi \coloneqq Fz^{[\pmb{d}]}$, where $z^{[\pmb{d}]}$ is defined as in (\ref{eqn:zAd}). With the transformation map
$
T(\eta,z) = \left(
\eta, 
z^{[\pmb{d}]}
 \right),
$
System (\ref{eqn:xyz}) is globally immersible into $\Pi(\begin{pmatrix}
A_{\eta} & F\\
\pmb{0} & A_z^{[\pmb{d}]}
\end{pmatrix}, \begin{pmatrix}
I_{n_{\eta} +n_z} & \pmb{0}
\end{pmatrix} )$. 
\end{theorem}
Proof: This is a direct consequence of the construction of the algebraic lifting above. $\Box$

\begin{remark}
It is worth noting that the algebraic lifting procedure can be also considered as Carleman linearization \cite{BOO:KS91}. In other words, for such systems, Carlement linearization is exact and finite.
\end{remark}

Nonlinear systems in the form of (\ref{eqn:xyz}) can often arise in the presence of a polynomial exogenous input generated from a Wiener system \cite{ART:NZ01,ART:TS14}, which consists of a linear dynamic model and a nonlinear output model due to nonlinear sensors. Consider the output regulation problem (see, e.g., Chapter 1 of \cite{BOO:KIF12}) of a linear system in the form of
$
\eta(t+1) = A_{\eta}\eta(t)+B_{\eta} u(t)+B_{v} \tilde{v}(t),
e(t+1) = C_{\eta} \eta(t) + C_v \bar{v}(t), ~~ t\in \mathbb{Z}^+,
$
where $\eta\in \mathbb{R}^{n_{\eta}}$ is the state, $u\in \mathbb{R}^{n_{u}}$ is the control input, $\tilde{v}$ and $\bar{v}$ are the exogenous inputs, which include disturbances (to be rejected) and/or references (to be tracked), $e$ is an error variable, and $A_{\eta},B_{\eta}$, $B_{v}$, $C_{\eta}$, and $C_v$ are some given matrices. The exogenous inputs $\tilde{v}$ and $\bar{v}$ are generated from a linear exogenous system given by
$
z(t+1) = A_z z(t), 
\tilde{v}(t) = \tilde{\varphi}(z(t)), ~\bar{v}(t) = \bar{\varphi}(z(t)), ~~t\in \mathbb{Z}^+,
$
where $z\in \mathbb{R}^{n_z}$ is the state of the generator, $\tilde{\varphi}(\cdot)$ and $\bar{\varphi}(\cdot)$ are polynomial functions, and $A_z$ is some given matrix. This extends the formulation of the output regulation problem in Chapter 1 of \cite{BOO:KIF12} to Wiener exogenous systems. Consider output regulation with full information (see \cite{BOO:KIF12} for details), the controller takes the form of 
$
u = Kx + \phi(z),
$
where $\phi(\cdot)$ is a polynomial function. Hence, the closed-loop system becomes
$
\eta(t+1) = (A_{\eta}+B_{\eta}K)\eta(t)+B_{\eta}\phi(z) + B_{v} \tilde{\varphi}(z),
z(t+1) = A_z z(t), t\in \mathbb{Z}^+ .
$

\subsection{Approximate immersion}\label{sec:appimmersion}
Since linear equivalents exist only for very particular classes of systems, we now introduce an approximate version of state immersion. In general cases, we want to find a transformation map $T:\mathbb{R}^n \rightarrow \mathbb{R}^{m}$ such that $T(f(x))-AT(x)$ is within some given tolerance for all $x$ inside some subset of $\mathbb{R}^n$. The formal definition of \emph{approximate immersions} is given below.

\begin{definition}\label{def:approximate}
Given a subset $S\subseteq \mathbb{R}^n$, a transformation map $T: \mathbb{R}^n \rightarrow \mathbb{R}^m$, matrices $A\in \mathbb{R}^{m\times m}, C\in \mathbb{R}^{n\times m}$ and a bounded set $\Delta \subset \mathbb{R}^{m}$,  System (\ref{eqn:fx}) is  \emph{$(S,\Delta)$-approximately immersible} into $\Pi(A,C)$ with $T(x)$ if 
$
\forall x\in S, T(f(x)) - AT(x) \in \Delta, CT(x) = x.
$
\end{definition}

To rigorously compute a bounded set of an \emph{approximate immersion}, the following definition is also needed.
\begin{definition}
Given a compact set $S \subset \mathbb{R}^n$ and a subset $\omega \subset S$, $\omega$ is called an $\epsilon$-covering of $S$ if $S\subseteq \omega+\epsilon \mathbb{B}_n$.
\end{definition}

We now present the construction of \emph{approximate immersions} in a specific subset of $X$. Since our goal is to compute an invariant set inside $X$, it is sufficient to consider \emph{approximate immersions} in the \textit{maximal admissible invariant set} $O_{\infty}$. However, what is available is $O_k$ for any $k\in \mathbb{Z}^+$ but not $O_{\infty}$ itself. For this reason, we define the following problem for any given $M\in \mathbb{Z}^+$:
\begin{subequations}\label{eqn:gMdelta}
\begin{align}
\delta_M^* &\coloneqq \min_{\delta,\pmb{\gamma}_M} \delta\\
\textrm{s.t.} &\|f^{M+1}(x)- \sum\limits_{\ell=0}^M \gamma_{\ell}f^{\ell}(x)\|_{\infty} \le \delta, \forall x\in O_{M+1}. \label{eqn:gMdeltaO}
\end{align}
\end{subequations}
where $\pmb{\gamma}_M\coloneqq   [
\gamma_{0} ~ \gamma_{1} ~ \cdots ~\gamma_{M}
]$.  An additional assumption is made to discuss the properties of Problem (\ref{eqn:gMdelta}).

(\textbf{A4}) For some $k\in \mathbb{Z}^+ $, there exists a class $\mathcal{K}$ function $c$ such that $\|f^t(x)-f^t(y)\|\le c(\|x-y\|)$ for all $t\in \mathbb{Z}^+$ and $x,y\in O_{k}$.

A sufficient condition for (\textbf{A4}) is that System (\ref{eqn:fx}) is incrementally stable (see, e.g., \cite{ART:A02} for the definition) in $O_{k}$ for some $k\in \mathbb{Z}^+$. However, this condition is more relaxed in the sense that it only requires that the distance of any two trajectories remains bounded by a class $\mathcal{K}$ function $c$ of the initial distance.  We then present properties of Problem (\ref{eqn:gMdelta}) in the following lemma.

\begin{lemma}\label{lem:delta}
Suppose (\textbf{A1}) $\&$ (\textbf{A2}) hold. Let $\delta_M^*$ be defined in (\ref{eqn:gMdelta}) for all $M\in \mathbb{Z}^+$.  Then, the following properties hold: (i) $\delta_{M+1}^*\le \delta_M^*$ for all $M\in \mathbb{Z}^+$. (ii) Moreover, $\lim_{M\rightarrow \infty} \delta_M^* = 0$ when (\textbf{A3}) $\&$ (\textbf{A4}) are also satisfied.
\end{lemma}
Proof: (i) For any $M\in \mathbb{Z}^+$, let $\delta_M^*$ be the optimal solution to Problem (\ref{eqn:gMdelta}) with $\pmb{\gamma}_M^* \coloneqq \{\gamma^*_{\ell}\in \mathbb{R}^{n\times n}\}_{\ell=0}^M$. Hence, $\|f^{M+1}(x)- \sum_{\ell=0}^M \gamma^*_{\ell}f^{\ell}(x)\|_{\infty} \le \delta_M^*, \forall x\in O_{M+1}$. From the definition of $\{O_k\}_{k\in \mathbb{Z}^+}$ in (\ref{eqn:Ok}), $x\in O_{M+2} \Rightarrow f(x) \in O_{M+1}$. Hence, for any $x\in O_{M+2}$, $\|f^{M+2}(x)- \sum_{\ell=0}^M \gamma^*_{\ell}f^{\ell+1}(x)\|_{\infty} = \|f^{M+1}(f(x))- \sum_{\ell=0}^M \gamma^*_{\ell}f^{\ell}(f(x))\|_{\infty} \le \delta_M^*$, which means $\delta_M^*$ is a feasible solution to Problem (\ref{eqn:gMdelta}) for $M+1$.  Thus, Property (i) holds. \\
(ii) Now we show that, for any $\epsilon>0$, there exists $M$ such that $\delta_M^*< \epsilon$ with (\textbf{A3}) $\&$ (\textbf{A4}). From Proposition \ref{prop:Oinfnl}, there exists $k^*$ such that $O_k = O_{\infty}$ for any $k\ge k^*$, which under (\textbf{A4}), implies that there exists a class $\mathcal{K}$ function $c$ such that $\|f^k(x)-f^k(y)\|\le c(\|x-y\|)$ for all $x,y\in O_{k^*}$. Given any $\varepsilon>0$, we can select $N$ points $\omega_N\coloneqq\{x_1,x_2,\cdots, x_N\}$ such that $\omega_N$ is a $\varepsilon$-covering of $O_{k^*}$ (or $O_{\infty}$). We consider the N-ary Cartesian power of $O_{\infty}$, denoted by $O_{\infty}^N \coloneqq \underbrace{O_{\infty} \times O_{\infty} \times \cdots \times O_{\infty}}_{N} \subset \mathbb{R}^{nN}$. Let us divide $O_{\infty}^N$ into disjoint subsets using a regular grid in which the diameter of each subset is less than $\varepsilon$. The number of the disjoint subsets is denoted by $N_{\varepsilon}$ (an upper bound can be easily obtained as $O_{\infty}^N$ is bounded). 
From the invariance of $O_{\infty}$, the stacked vector $(f^k(x_1),f^k(x_2),\cdots, f^k(x_N))$ is contained in $O_{\infty}^N$ for any $k\in \mathbb{Z}^+$. Then, the pigeonhole principle suggests that at least two points in $\{(f^k(x_1),f^k(x_2),\cdots, f^k(x_N))\}_{k=0}^{N_{\varepsilon}}$ fall into the same subset, say $(f^{k_1}(x_1),f^{k_1}(x_2),\cdots, f^{k_1}(x_N))$ and $(f^{k_2}(x_1),f^{k_2}(x_2),\cdots, f^{k_2}(x_N))$ with $k_2 > k_1\ge k^*$. Hence, $\|f^{k_2}(x)-f^{k_1}(x)\|_{\infty} \le \|f^{k_2}(x)-f^{k_1}(x)\|_{2} \le \varepsilon $ for any $x\in \omega_N$. Since $\omega_N$ is a $\varepsilon$-covering of $O_{\infty}$, for any $x\in O_{\infty}$, there exists $x'\in \omega_N$ such that $\|x-x'\|\le \varepsilon$, which implies that $\|f^{k_2}(x)-f^{k_1}(x)\|_{\infty} \le \|f^{k_2}(x)-f^{k_1}(x)\|_{2} = \|f^{k_2}(x)-f^{k_2}(x')\|+\|f^{k_1}(x')-f^{k_1}(x)\|+\|f^{k_2}(x')-f^{k_1}(x')\|_{2} \le c(\varepsilon) + c(\varepsilon) + \varepsilon$, where the last inequality follows from \textbf{A4}. Let $\varepsilon$ be chosen such that $2c(\varepsilon)+\varepsilon = \epsilon$. Then, we conclude that $\delta_{k_2-1}^* \le \epsilon$. Hence, from (i), $\delta_{M}^* \le \epsilon$ for any $M \ge k_2-1$. This completes the proof.   $\Box$

From the convergence of $\{\delta_M^*\}_{M\in \mathbb{Z}^+}$, the following statement can be made.

\begin{theorem}\label{thm:approximate}
Suppose (\textbf{A1})--(\textbf{A4}) hold. Let $O_{\infty}$ be defined as in (\ref{eqn:Oinf}). For any given $\delta > 0$, there exist a finite $m\in \mathbb{Z}^+$, a continuous \emph{linearly independent} map $T: \mathbb{R}^n \rightarrow \mathbb{R}^{m}$, an observable pair $(C,A)$ with $C\in \mathbb{R}^{n\times m}$ and $A\in \mathbb{R}^{m\times m}$, and a matrix $B\in \mathbb{R}^{m\times n}$ such that System (\ref{eqn:fx}) is \emph{$(O_{\infty},B\Delta_{\delta})$-approximately immersible} into $\Pi(A,C)$ with $T(x)$, where
\begin{align}\label{eqn:Wdelta}
\Delta_{\delta}\coloneqq\{v\in \mathbb{R}^n:\|v\|_{\infty}\le \delta\}.
\end{align}
\end{theorem}
Proof: From Lemma \ref{lem:delta}, for any $\delta >0$, there always exist $M\in \mathbb{Z}^+$ and matrices $\{\gamma_{\ell}\in \mathbb{R}^{n\times n}\}_{\ell=0}^M$ such that (\ref{eqn:gMdeltaO}) holds. Hence,
$
\mathcal{F}_M(f(x))
- \Gamma(\pmb{\gamma}_M)
\mathcal{F}_M(x)
\in 
\left( \begin{array}{c}
\pmb{0}_{Mn\times n}\\
I_n
\end{array} \right) \Delta_{\delta}
$
for all $x\in O_{\infty}$. Let $T(x)$ be the $m$ \emph{linearly independent} functions that form a basis for the spanning set of $\{x_1, x_2, \cdots, f^M_n(x)\}$ in $O_{\infty}$. We can find a full column rank matrix $P\in \mathbb{R}^{(M+1)n\times m}$ such that
$
T(f(x))- P^+\Gamma(\pmb{\gamma}_M) PT(x) 
\in  P^+
\left( \begin{array}{c}
\pmb{0}_{Mn\times n}\\
I_n
\end{array} \right) \Delta_{\delta}, \forall x\in O_{\infty}.
$
Letting $A = P^+\Gamma(\pmb{\gamma}_M) P, C = [I_n ~ \pmb{0}_{n\times (M+1)n}]P$ and $B = P^+
\left( \begin{array}{c}
\pmb{0}_{Mn\times n}\\
I_n
\end{array} \right)$ yields the approximate immersibility property. When $(C,A)$ is observable, the statement holds. Otherwise, we consider the observable subspace and get a new pair of $C$ and $A$. This completes the proof. $\Box$

In practice, it is not realistic to solve Problem (\ref{eqn:gMdelta}) exactly as there are infinitely many constraints. Instead, we solve a sampled problem with a finite sample. For any $M\in \mathbb{R}^{n}$, given a sample $\omega \subset O_{M+1}$, the following sampled problem is defined
\begin{subequations}\label{eqn:gMdeltas}
\begin{align}
\delta_M(\omega) &\coloneqq \min_{\delta,\pmb{\gamma}_M} \delta\\
\textrm{s.t.} &\|f^{M+1}(x)- \sum\limits_{\ell=0}^M \gamma_{\ell}f^{\ell}(x)\|_{\infty} \le \delta, \forall x\in \omega. \label{eqn:gMdeltaOs}
\end{align}
\end{subequations}
When the sample $\omega$ is an $\epsilon$-covering of $O_{M+1}$ for some $\epsilon>0$, an \emph{approximate immersion} can be obtained, as stated in the following proposition.

\begin{proposition}\label{prop:mis}
Suppose (\textbf{A1}) $\&$ (\textbf{A2}) hold. For any $k\in \mathbb{Z}^+$, let us define $O_k$ as in (\ref{eqn:Ok}) with $O_{\infty}$ being as in (\ref{eqn:Oinf}). Given any $M\in \mathbb{Z}^+$ and an $\epsilon$-covering of $O_{M+1}$ for some $\epsilon>0$, denoted by $\omega$, let $\delta_M(\omega)$ be defined as in (\ref{eqn:gMdeltas}). Then, there exist a Lipschtiz continuous map $T: \mathbb{R}^n \rightarrow \mathbb{R}^{m}$ in $O_{\infty}$ with the Lipschitz constant $L_T>0$ and matrices $(A,B,C)$ such that $\Pi(A,C)$ is $(O_{\infty},B \Delta_{\delta_M(\omega)} +  L_T(L_f +\|A\|)\epsilon \mathbb{B}_m)$-approximately immersible to System (\ref{eqn:fx}), where $\Delta_{\delta_M(\omega)}$ is given as in (\ref{eqn:Wdelta}).
\end{proposition}
Proof: Following the same arguments in the proof of Theorem \ref{thm:approximate}, there exist $T(x)$ and matrices $(A,B,C)$ such that $T(f(x))-AT(x) \in B \Delta_{\delta_M(\omega)}$ for all $x\in \omega$, where $T(x)$ is chosen to be a subset of $\{x_1, x_2, \cdots, f^M_n(x)\}$. From the Lipschitz continuity of $f(x)$ in $X$ with the Lipschitz constant in (\textbf{A1}), a Lipschitz constant of $T(x)$ can be easily obtained, denoted by $L_T$. Since $\omega$ is an $\epsilon$-covering of $O_{M+1}$, for any $x\in O_{M+1} $, there exists $x'\in \omega$ such that $\|x-x'\|\le \epsilon$, which implies that $T(f(x))-AT(x) \in  T(f(x'))-AT(x') + L_T(L_f +\|A\|)\epsilon \mathbb{B}_m \in B \Delta_{\delta_M(\omega)} +  L_T(L_f +\|A\|)\epsilon \mathbb{B}_m$. This completes the proof. $\Box$

\subsection{Sampling and finite-sample guarantees}\label{sec:sampling}
In the rest of this section, we discuss the sampling procedure in (\ref{eqn:gMdeltas}) with formal guarantees. We first select a raw data set on $X$ (or a box enclosing $X$), denoted as $\mathcal{D}$, and generate a trajectory with a sufficiently long horizon from each point. Given $k\in \mathbb{Z}^+$, to sample points on the set $O_k$, we then pick the points inside $O_k$ as follows: $\omega=\mathcal{D} \cap O_k = \{x\in \mathcal{D}: f^\ell(x)\in X, \ell=0,1,\cdots, k\}$. More details of this sampling procedure can be found in \cite{ART:WJ20a}. However, in general, without regularity conditions, it is difficult to estimate a compact set from sampled points, see, e.g., \cite{ART:CR04}. In this paper, we tackle this problem by enlarging the sampling region. To this end, we extend the continuity condition in \textbf{A1} as follows: (\textbf{A1}') The function $f(x)$ is Lipschitz continuous in $X+\rho\mathbb{B}_n$ with a Lipschitz constant $L_f$ for some $\rho>0$.

To derive formal guarantees on $\epsilon$-covering, we need the following lemma. 
\begin{lemma}\label{lem:Oepsilon}
Given some $\rho>0$, suppose \textbf{A1}' holds. Let $\{O_k\}_{k\in \mathbb{Z}^+}$ and $\{O^{\rho}_k\}_{k\in \mathbb{Z}^+}$ be defined by the iteration in (\ref{eqn:Ok}) with $O_0 = X$ and $O^{\rho}_0 = X+\rho\mathbb{B}_n$. Then, it holds that
\begin{align}
O_k + \min\{1,\frac{1}{L_f^k}\}\rho \mathbb{B}_n \subseteq O^{\rho}_k, \forall k\in \mathbb{Z}^+.
\end{align}
\end{lemma}
\vspace{-8mm}
Proof: From the iteration in (\ref{eqn:Ok}), it can be verified that $O^{\rho}_k=\{x\in \mathbb{R}^n: f^\ell(x) \in X+\rho \mathbb{B}_n,  \ell=0,1,\cdots, k\}$. For any $k\in \mathbb{Z}^+$, we consider any $x\in O_k + \min\{1,\frac{1}{L_f^k}\}\rho \mathbb{B}_n$, which can be expressed as $x=y+\min\{1,\frac{1}{L_f^k}\}\rho s$ for some $y\in O_k$ and $s\in \mathbb{B}_n$. We can show recursively that $f^\ell(x) \in f^\ell(y)+L_f^\ell\min\{1,\frac{1}{L_f^k}\}\rho \mathbb{B}_n \subseteq X+\rho \mathbb{B}_n$ for any $\ell=0,1,\cdots, k$ from the fact that $L_f^\ell\min\{1,\frac{1}{L_f^k}\} \le 1$ for any $\ell \le k$. Hence, $x\in O^{\rho}_k$. This completes the proof. $\Box$

Let us also recall the definition of $\epsilon$-packing of a set, see, e.g.,  \cite[Chapter 27]{BOO:SB14}.
\begin{definition}\label{def:packing}
Given a compact set $S\subset \mathbb{R}^n$ and $\epsilon>0$,  a (finite) subset $P$ of $S$ is called an $\epsilon$-packing if $\min_{x\in P,y\in P, x\not=y} \|x-y\|>\epsilon$. The packing number, denoted by $\mathcal{N}(S,\epsilon)$, is the maximal cardinality of any $\epsilon$-packing of $S$.
\end{definition}

We first consider the case that the sample in (\ref{eqn:gMdeltas}) is obtained from a uniform grid defined below: 
\begin{align}\label{eqn:Gn}
\eta \mathbb{G}^n:= \{c\in \mathbb{R}^n: c_i = k_i \eta, k_i\in \mathbb{Z},\forall i\}
\end{align}
where $\eta \in \mathbb{R}^+$ is the grid parameter. We then claim that the sample obtained from a fine grid is a concrete $\epsilon$-covering, as stated in the following proposition.

\begin{proposition}\label{prop:grid}
Consider the same conditions in Lemma \ref{lem:Oepsilon}. For any $\eta \in \mathbb{R}^+$, let $\eta \mathbb{G}^n$ be defined as in (\ref{eqn:Gn}). Given any $k\in \mathbb{Z}^+$, if $\eta \le \min\{1,\frac{1}{L_f^k}\}\rho$, $\eta \mathbb{G}^n \cap O^{\rho}_k$ is a $\eta \sqrt{n}  $-covering of $O_k$.
\end{proposition}
Proof: For any $x\in \mathbb{R}^n$, it can be verified that $\min_{c\in \eta \mathbb{G}^n }\|x-c\|_{2}\le \sqrt{n}\min_{c\in \eta \mathbb{G}^n }\|x-c\|_{\infty} \le \frac{\eta \sqrt{n}}{2}
$. Hence, $x+\frac{\eta \sqrt{n}}{2} \mathbb{B}_n \cap \eta \mathbb{G}^n \not=\emptyset$ for any $x\in \mathbb{R}^n$. For any $k\in \mathbb{Z}^+$,  we consider the maximal $\frac{\eta \sqrt{n}}{2}$-packing of $O_k$, denoted by $P_{\eta}$. From Definition \ref{def:packing}, $P_{\eta}$ is also a $\frac{\eta \sqrt{n}}{2}$-covering of $O_k$. Suppose $\eta \le \frac{2}{\sqrt{n}}\min\{1,\frac{1}{L_f^k}\}\rho$, we have that $O_k \subseteq P_{\eta} + \frac{\eta \sqrt{n}}{2} \mathbb{B}_n \subseteq O_k + \frac{\eta \sqrt{n}}{2} \mathbb{B}_n \subseteq  O_k^\rho $ where the last inclusion is from Lemma \ref{lem:Oepsilon}. Since, for any $x\in P_{\eta}$, $x+\frac{\eta \sqrt{n}}{2} \mathbb{B}_n$ contains at least one point in $\eta \mathbb{G}^n \cap O^{\rho}_k$, $P_{\eta} \subseteq \eta \mathbb{G}^n \cap O^{\rho}_k + \frac{\eta \sqrt{n}}{2} \mathbb{B}_n$. Therefore, $O_k \subseteq P_{\eta} + \frac{\eta \sqrt{n}}{2} \mathbb{B}_n \subseteq \mathbb{G}^n \cap O^{\rho}_k + \eta \sqrt{n} \mathbb{B}_n$. This completes the proof. $\Box$

We also consider the case of random sampling, in which probabilistic guarantees on $\epsilon$-covering can be derived.

\begin{proposition}\label{prop:random}
Consider the same conditions in Lemma \ref{lem:Oepsilon}. Given any $k\in \mathbb{Z}^+$, suppose $\omega$ is independent and identically distributed (i.i.d.) with respective to the uniform distribution over $O_k^{\rho}$ with $|\omega|=N$, then, for any $\eta \le \min\{1,\frac{1}{L_f^k}\}\rho$, with probability no smaller than $1-\mathcal{N}(O_{k},\eta) (1-\frac{\textrm{vol}(\eta \mathbb{B}_n)}{\textrm{vol}(X+\rho \mathbb{B}_n)})^N$, $\omega$ is a $2\eta$-covering of $O_{k}$, where $\textrm{vol}(\cdot)$ denotes the volume and $\mathcal{N}(O_{k},\eta)$ is given in Definition \ref{def:packing}.
\end{proposition}
Proof: For any $\eta \le \min\{1,\frac{1}{L_f^k}\}\rho$, we consider the maximal $\eta$-packing of $O_k$, denoted by $P_{\eta}$. From Lemma \ref{lem:Oepsilon}, we know that $O_k+\eta \mathbb{B}_n \subseteq O_{k}^\rho$. The probability that $x+\eta \mathbb{B}_n \cap \omega = \emptyset$ is $(1-\frac{\textrm{vol}(\eta \mathbb{B}_n)}{\textrm{vol}(O_k^{\rho})})^N \le (1-\frac{\textrm{vol}(\eta \mathbb{B}_n)}{\textrm{vol}(X+\rho \mathbb{B}_n)})^N$ for any $x\in P_{\eta}$. Hence, the probability that $x+\eta \mathbb{B}_n \cap \omega \not= \emptyset$ for any $x\in  P_{\eta}$ is no smaller than $1-\mathcal{N}(O_{k},\eta) (1-\frac{\textrm{vol}(\eta \mathbb{B}_n)}{\textrm{vol}(X+\rho \mathbb{B}_n)})^N$ (as $|P_{\eta}| = \mathcal{N}(O_{k},\eta)$ from the definition). Finally, since  $x+\eta \mathbb{B}_n \cap \omega \not= \emptyset$ for any $x\in P_{\eta}$ implies that $P_{\eta}+\eta \mathbb{B}_n \subseteq \omega + 2\eta \mathbb{B}_n$, we conclude the statement. $\Box$

\begin{remark}
For any $k\in \mathbb{Z}^+$ and any $\eta \in \mathbb{R}^+$, $\mathcal{N}(O_{k},\eta)$ can be bounded from above as
$
\mathcal{N}(O_k,\eta) \le \frac{\textrm{vol}(O_k+\frac{\eta}{2}\mathbb{B}_n)}{\textrm{vol}(\frac{\eta}{2}\mathbb{B}_n)}\le \frac{\textrm{vol}(X+\frac{\eta}{2}\mathbb{B}_n)}{\textrm{vol}(\frac{\eta}{2}\mathbb{B}_n)}.
$
\end{remark}

It is worth noting that there also exist asymptotic probabilistic bounds for random covering problems, see, e.g., \cite{ART:J86}. However, these bounds are not applicable because we consider a finite number of points. With the discussions above, we also want to mention that the result in Proposition \ref{prop:mis} remains the same with the enlarged sampling region except that $\omega$ is contained in $O_{M+1}^{\rho}$ but not necessarily in $O_{M+1}$ and the Lipschitz constant $L_T$ of the map $T(x)$ is valid in $O_{M}^{\rho}$.

\section{Invariant set computation via immersion}\label{sec:invimm}
In this section, based on the discussion on immersion, we present the proposed immersion-based approach for computing invariant sets of nonlinear systems.

\subsection{Set invariance under immersion}
For systems that are immersible into a linear system (see Definition \ref{def:imm}),  we can also establish the immersion on invariant sets of the nonlinear system and its linear equivalent, as shown in the following proposition.
\begin{proposition}\label{prop:invariance}
Given the constraint set $X\subseteq \mathbb{R}^n$, suppose there exist a continuous map $T: \mathbb{R}^n \rightarrow \mathbb{R}^{m}$ in $X$ and matrices $A \in \mathbb{R}^{m\times m}, C\in \mathbb{R}^{n\times m}$ such that $AT(x) = T(f(x))$ and $CT(x) = x$ for all $x\in X$. Let $Z\subseteq X$ be an invariant set for System (\ref{eqn:fx}) and $\Xi \subseteq  \{\xi\in \mathbb{R}^m:C\xi \in X\}$ be an invariant set for $\Pi(A,C)$. Then,
(i) $T^{-1}(\Xi)\coloneqq\{x\in \mathbb{R}^n: T(x)\in \Xi\}\subseteq X$ is invariant for System (\ref{eqn:fx});
(ii) $T(Z)\subseteq \{\xi\in \mathbb{R}^m:C\xi \in X\}$ is invariant for $\Pi(A,C)$.
\end{proposition}
Proof: The proof can be found in Proposition 3 of \cite{ART:WJO20}. It is not repeated due to page limitation. $\Box$

The results in Proposition \ref{prop:invariance} allow to use the lifted linear system to compute the \emph{maximal admissible invariant set} $O_{\infty}$ of the system (\ref{eqn:fx}). Once the  \emph{maximal admissible invariant set} of the lifted linear system is computed, a closed-form (nonlinear) expression of $O_{\infty}$ can be obtained.  Given any pair $(A,C)$ with $A\in \mathbb{R}^{m\times m}$ and $C\in \mathbb{R}^{n\times m}$, let us define
\begin{align}\label{eqn:OinfL}
O_{\infty}^L(A,C) \coloneqq\{x\in \mathbb{R}^{m}: CA^{k}x\in X, \forall k\in \mathbb{Z}^+\}
\end{align}
From Theorem 4.1 in \cite{ART:GT91}, $O_{\infty}^L(A,C)$ exists and can be finitely determined when $(C,A)$ is observable and $A$ is Schur stable. From the understanding on set invariance under immersion, the following theorem can be obtained.

\begin{theorem}\label{thm:Omax}
Suppose (\textbf{A1}) $\&$ (\textbf{A2}) hold, let $O_{\infty}$ be defined as in (\ref{eqn:Oinf}) for System (\ref{eqn:fx}) with the constraint set $X$. Assume that System (\ref{eqn:fx}) is immersible into a linear system $\Pi(A_{\xi},C_{\xi})$ in (\ref{eqn:xiy2}). Then, there exist a continuous \emph{linearly independent} map $T:\mathbb{R}^n \rightarrow \mathbb{R}^m$ in $O_{\infty}$ and an observable pair $(C,A)$ such that $O_{\infty}^L(A,C)$ is compact and $O_{\infty}=T^{-1}(O_{\infty}^L(A,C))$, where $O_{\infty}^L(A,C)$ is defined in (\ref{eqn:OinfL}).
\end{theorem}

Proof: From Theorem \ref{thm:immT}, there always exist a continuous \emph{linearly independent} map $T:\mathbb{R}^n \rightarrow \mathbb{R}^m$ in $O_{\infty}$ and an observable pair $(C,A)$ such that $AT(x) = T(f(x))$ and $CT(x) = x$ for any $x\in O_{\infty}$. From Theorem 2.1 in \cite{ART:GT91}, $O_{\infty}^L(A,C)$ is compact as $(C,A)$ is observable. Now, we need to show that $O_{\infty}=T^{-1}(O_{\infty}^L(A,C))$. From Proposition \ref{prop:invariance}, $T(O_{\infty})\subseteq \{\xi\in \mathbb{R}^m:C\xi \in X\}$ is invariant for $\Pi(A,C)$ and $T^{-1}(O_{\infty}^L(A,C))\subseteq X$ is invariant for System (\ref{eqn:fx}). Since $O_{\infty}$ and $O_{\infty}^L(A,C)$ are the \emph{maximal admissible invariant sets} for System (\ref{eqn:fx}) and $\Pi(A,C)$ respectively, $T(O_{\infty}) \subseteq O_{\infty}^L(A,C)$ and $T^{-1}(O_{\infty}^L(A,C)) \subseteq O_{\infty}$, which implies that $O_{\infty}=T^{-1}(O_{\infty}^L(A,C))$. 
$\Box$

%\begin{remark}
%The purpose of \textbf{A1} is to guarantee that the lifted linear system is asymptotically stable, which ensures the existence of $O_{\infty}^L(A,C)$. However, as shown in \cite{ART:GT91}, asymptotic stability is not a necessary condition for the existence of the \emph{maximal invariant set}. Hence, \textbf{A1} may not be needed as long as the \emph{maximal invariant set} of the lifted linear system can be efficiently computed. 
%\end{remark}

\subsection{An inner approximation}
As mentioned in Section \ref{sec:appimmersion}, for general nonlinear systems, we can only achieve \emph{approximate immersions}. Suppose a $(O_{\infty},\Delta)$-approximately immersion $\Pi(A,C)$ is available with some transformation map $T(x)$, as defined in Definition \ref{def:approximate}. To account for the mismatch between System (\ref{eqn:fx}) and the linear system $\Pi(A,C)$, we compute a tightened subset of  $O_{\infty}^L(A,C)$, instead of $O_{\infty}^L(A,C)$. Given $(A,C)$ and $\Delta $, let us define
\begin{align}\label{eqn:OinfLdelta}
O_{\infty}^{L,\Delta}&(A,C) \coloneqq\{x\in \mathbb{R}^{m}: \nonumber\\
 &CA^{k}x\in X\ominus \sum_{\ell=0}^{k-1}CA^{\ell}\Delta, \forall k\in \mathbb{Z}^+\}
\end{align}
From \cite{ART:KG98}, the set $O_{\infty}^{L,\Delta}(A,C)$ is nonempty when $\sum_{\ell=0}^{\infty}CA^{\ell}\Delta \subseteq X$ and it is the \emph{maximal admissible robust invariant set} for the disturbed system $x^+ = Ax+w$ where the disturbance $w$ is constrained in $\Delta$. From the set defined in (\ref{eqn:OinfLdelta}), an inner approximation of $\O_{\infty}$ can be obtained, as stated in the following theorem.

\begin{theorem}\label{thm:innerset}
Suppose (\textbf{A1}) $\&$ (\textbf{A2}) hold. Let $O_{\infty}$ be defined as in (\ref{eqn:Oinf}) for System (\ref{eqn:fx}). Consider a continuous \emph{linearly independent} map $T:\mathbb{R}^n\rightarrow \mathbb{R}^m$ in $O_{\infty}$, an observable pair $(C,A)$ and $\Delta \subset \mathbb{R}^m$ such that System (\ref{eqn:fx}) is $(O_{\infty},\Delta)$-approximately immersible to $\Pi(A,C)$, the following results hold: (i) $T^{-1}(O_{\infty}^{L,\Delta}(A,C))\subseteq O_{\infty}$;  (ii) $T^{-1}(O_{\infty}^{L,\Delta}(A,C))$ is invariant for System (\ref{eqn:fx}), where $O_{\infty}^{L,\Delta}(A,C)$ is defined as in (\ref{eqn:OinfLdelta}).
\end{theorem}
Proof: (i) First, we show that $T^{-1}(O_{\infty}^{L,\Delta}(A,C))\subseteq O_{\infty}$. For any $x\in T^{-1}(O_{\infty}^{L,\Delta}(A,C))$, we know that $CA^kT(x)\in X\ominus \sum_{\ell=0}^{k-1}CA^{\ell}\Delta$ for all $k\in \mathbb{Z}^+$. Since $CT(x)=x$, it is obvious that $x\in X$. From the fact that $T(f(x))-AT(x)\in \Delta$, we know that 
$
f(x) = CT(f(x))\in CAT(x)+C\Delta \subseteq X\ominus C\Delta + C\Delta \subseteq X
$, where the last inclusion follows from the properties of the Minkowski difference, see, e.g., Theorem 2.1 in \cite{ART:KG98}. Hence, it holds that $x\in O_1$. The proof goes by induction. Suppose $x\in O_k$ for some $k\in \mathbb{Z}^+$. We can see that $T(f(f^\ell(x)))-AT(f^\ell(x))\in B\Delta$ for all $\ell =0,1,\cdots, k$ because $f^\ell(x)\in X$. Hence, 
$
f^{k+1}(x)=CT(f^{k+1}(x))\in CAT(f^k(x))+C\Delta
\subseteq CA^2T(f^{k-1}(x))+CA\Delta +C\Delta \subseteq \cdots 
%&\subseteq CA^{k+1}T(x) + \sum_{\ell=0}^{k}CA^{\ell}\Delta \\
\subseteq X\ominus \sum_{\ell=0}^{k}CA^{\ell}\Delta + \sum_{\ell=0}^{k}CA^{\ell}\Delta \subseteq X
$
This implies that $x\in O_{k+1}$. Therefore, we conclude that $x\in O_{\infty}$.  (ii) To prove the invariance of $T^{-1}(O_{\infty}^{L,\Delta}(A,C))$, we need to show that $f(x)\in T^{-1}(O_{\infty}^{L,\Delta}(A,C))$, which means that $CA^kT(f(x))\in X\ominus \sum_{\ell=0}^{k-1}CA^{\ell}\Delta$ for all $k\in \mathbb{Z}^+$. Since $T(f(x)) \in AT(x)+\Delta$, it holds that
$
CA^kT(f(x))\in CA^{k+1}T(x)+CA^k\Delta 
\subseteq  X\ominus \sum_{\ell=0}^{k}CA^{\ell}\Delta + CA^k\Delta
\subseteq X\ominus \sum_{\ell=0}^{k-1}CA^{\ell}\Delta,
$
for any $k\in \mathbb{Z}^+$.
$\Box$

\begin{remark}
In the presence of additive (bounded) disturbances in System (\ref{eqn:fx}), we also have to take both the mismatch error and the disturbances into consideration in the tighteed set in (\ref{eqn:OinfLdelta}).
\end{remark}

\section{Computational aspects}\label{sec:com}
This section discusses some computational aspects of the proposed method.

\subsection{Numerical solution for approximate immersion}
To characterize invariant sets, we first need to compute an \emph{approximate immersion} with a mismatch bound. As shown in Section \ref{sec:appimmersion}, this can be done by solving Problem (\ref{eqn:gMdelta}). However, this problem has infinite number of constraints. For this reason, we solve the sampled problem (\ref{eqn:gMdeltas}). We follow the sampling procedure in Section \ref{sec:sampling} to get a sample $\omega$ for some given $M\in \mathbb{Z}^+$ and  formulate Problem (\ref{eqn:gMdeltas}). As $|\omega|$ is usually quite large, it is expensive to solve Problem (\ref{eqn:gMdeltas}) exactly.  Instead, for numerical efficiency, we solve the following least squares regression problem,
\begin{align}\label{eqn:sumsumgammaM}
\min_{\pmb{\gamma}_M} \sum_{x \in \omega} \|f^{M+1}(x)-  \pmb{\gamma}_M \mathcal{F}_M(x)\|^2_{2} 
\end{align}
where $\mathcal{F}_M(x)$ is defined as in (\ref{eqn:FMx}).  For numerical stability, a regularized problem is solved. Let the solution of Problem (\ref{eqn:sumsumgammaM}) be denoted by $\hat{\pmb{\gamma}}_M$. With this solution, we can compute
\begin{align}\label{eqn:hatdeltaM}
\hat{\delta}_M = \max_{x \in \omega_N \cap O_{M+1}}  \|f^{M+1}(x)- \sum\limits_{\ell=0}^M \hat{\gamma}_{\ell}f^{\ell}(x)\|_{\infty}.
\end{align}
From $\hat{\pmb{\gamma}}_{M}$, a linear system $\Pi(\Gamma(\hat{\pmb{\gamma}}_{M}),[I_n~\pmb{0}_{n\times (M+1)n}])$ can be obtained with the transformation $\mathcal{F}_M(x)$. By checking and removing the redundancy, we get a \emph{linearly independent} transformation map $T: \mathbb{R}^n \rightarrow \mathbb{R}^m$ in $\omega_N \cap O_{M+1}$ and a full column rank matrix $P\in \mathbb{R}^{(M+1)n\times m}$ such that
$
\mathcal{F}_M(x)= PT(x), \forall x, 
$
which implies that
\begin{align}\label{eqn:TxP}
T(x) = P^+ \mathcal{F}_M(x)
\end{align}
where $P^+$ denotes the pseudo inverse of $P$. Then, we can get a linear system $\Pi(A_M,C_M)$ with $A_M=P^+ \Gamma(\pmb{\gamma}_{M})P$ and $C_M = [I_n~\pmb{0}_{n\times (M+1)n}]P$, and a matrix $B_M = P^+
\left( \begin{array}{c}
\pmb{0}_{Mn\times n}\\
I_n
\end{array} \right)$. Then, let $\hat{\Delta}_M = B_M \Delta_{\hat{\delta}_M}$ be the bound on the mismatch. Note that $
\mathcal{F}_M(x)
$ is already \emph{linearly independent} in many real applications.

\begin{remark}
With Propositions \ref{prop:grid} \& \ref{prop:random} in Section \ref{sec:sampling}, we can also compute a concrete mismatch error with the Lipschitz constant $L_f$.
\end{remark}

%\begin{align}
%\inf_{S\succ 0,U^\top U = I,\Sigma \succeq 0, \|\Sigma\|\le 1} \frac{1}{2} \|\mathcal{X}^+ - S^{-1}U\Sigma S \mathcal{X}\|
%\end{align}

\subsection{Computing the invariant set}

From the computations above, we can obtain an approximate mismatch bound $\hat{\Delta}_M$ and the linearized system $\Pi(A_{M},C_{M})$ for the given $M\in \mathbb{Z}^+$. When $\hat{\Delta}_M$ is sufficiently small, we compute $O_{\infty}^{L,\hat{\Delta}_M}(A_{M},C_{M})$ using the standard fixed-point algorithm \cite{ART:KG98}. Let 
\begin{align}\label{eqn:Omegadelta}
\Omega_{M}\coloneqq O_{\infty}^{L,\hat{\Delta}_M}(A_{M},C_{M}).
\end{align}
If $\Omega_{M}$ is empty, we will have to increase $M$ and repeat the computations above again. After a non-empty $\Omega_{M}$ is obtained, we can immediately compute its preimage $T^{-1}(\Omega_{M})$. The overall procedure is summarized in the following algorithm. 

\begin{algorithm}[h]
\caption{Invariant set computation via immersion}
\begin{algorithmic}[1]
\renewcommand{\algorithmicrequire}{\textbf{Input:}}
\renewcommand{\algorithmicensure}{\textbf{Output:}}
\REQUIRE $f(x)$, $X$, $\delta>0$, and $t_f$
\ENSURE $M$, $\Omega_{M}$ and $T(x)$\\
\textit{Initialization}:  Set $M \leftarrow 0$, take $N$ points $\omega_N$ inside $X$ by gridding (or random sampling) and generate the trajectory with the horizon $t_f$ for each point;
\STATE Solve Problem (\ref{eqn:sumsumgammaM}) and obtain $\hat{\pmb{\gamma}}_M$;
\STATE Compute $\hat{\delta}_M$ from (\ref{eqn:hatdeltaM}); 
\IF{$\hat{\delta}_M<\delta$}
\STATE Obtain $T(x)$ from (\ref{eqn:TxP});
\STATE Let $A_M \leftarrow P^+ \Gamma(\pmb{\gamma}_{M})P$, $C_M \leftarrow [I_n~\pmb{0}_{n\times (M+1)n}]P$, $B_M \leftarrow P^+
\left( \begin{array}{c}
\pmb{0}_{Mn\times n}\\
I_n
\end{array} \right)$, and $\hat{\Delta}_M = B_M \Delta_{\hat{\delta}_M}$;
\STATE Let $\hat{M} \leftarrow M$ and compute $\Omega_{M}$ defined in (\ref{eqn:Omegadelta});
\IF{$\Omega_{M}$ is empty}
\STATE Reduce the given $\delta$, set $M \leftarrow M+1$ and return to Step 1;
\ELSE 
\STATE Terminate and return $\Omega_{M}$ and $T(x)$.
\ENDIF
\ELSE 
\STATE Set $M \leftarrow M+1$ and return to Step 1;
\ENDIF
\end{algorithmic}
\label{algo:datadriven}
\end{algorithm}

\section{Numerical examples}\label{sec:num}

\begin{figure}[h]
\centering
\includegraphics[width=0.7\linewidth]{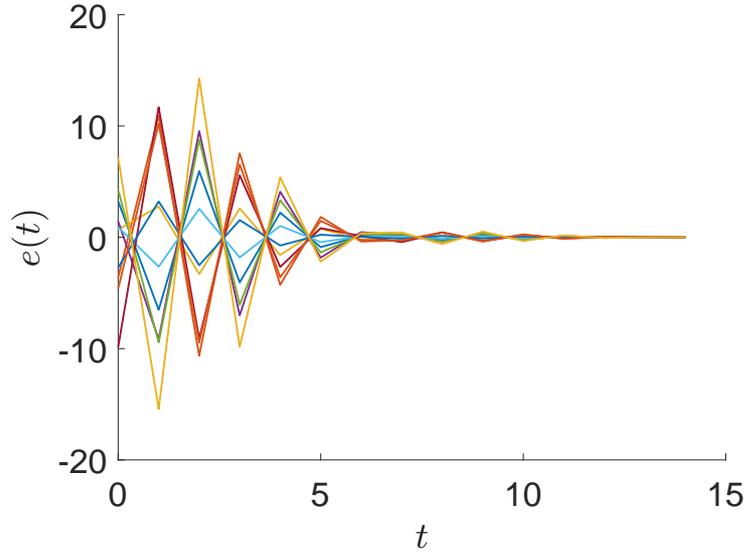}
\caption{Curves of output regulation error with different initial states.} 
\label{fig:outputerror}
\end{figure}

\textbf{Example 1} We first show an example in which the immersibility property holds exactly as described in Definition \ref{def:imm}. We consider output regulation (see Chapter 1 of \cite{BOO:KIF12}) of polynomial exogenous input. The plant is a linear system given as follows:
$
\eta^+ = A_{\eta} \eta +Bu, 
e =C_{\eta}\eta - v, 
$
where $A_{\eta} = [
1.1 ~ 1;
0 ~ 1.3
]$, $B = [
1;
1
]$, $C_{\eta} = [
1 ~ 0
]$ and $v$ is an exogenous input from an exosystem given by: 
$
z^+ = A_z z, y = C_z z, v = y + 0.3y^2-0.5y^3
$
where $A_z = [
0.6 ~ 0.8;
-0.8 ~ 0.6
]$ and $C_z = [
1 ~ -1
]$. The plant is controllable and the exosystem is marginally stable (the spectral radius is $1$). As mentioned in Section \ref{sec:special}, this exosystem can be called a Wiener system. Based on the algebraic lifting in (\ref{eqn:zAd}) and the internal model principle condition in Theorem 1.3.1 of \cite{BOO:KIF12}, output regulation can be achieved using  the following full-state feedback controller
$
u = K x + L_1 z^{[1]} + L_2 z^{[2]} + L_3 z^{[3]}
$
where $K = [-3.5 ~ 0]$, $L_1 = [2.2521~-2.4055], L_2 =[ 1.1677~-0.3187~0.9580],$ and $L_3 = [ -1.7681 ~ 1.0619 ~ -0.8383 ~1.7924]$. Thus, the closed-loop system is in the form of (\ref{eqn:xyz}) with a polynomial function of degree $3$. To verify this closed-loop system, we randomly generate several initial states and the curves of output regulation error $e$ are shown in Figure \ref{fig:outputerror}.

\begin{figure}[h]
\centering
  \begin{tabular}{cc}
  \subcaptionbox{$\mathcal{P}_{\eta}(O_{\infty})$ \label{fig:numtest}}{\includegraphics[width=0.48\linewidth]{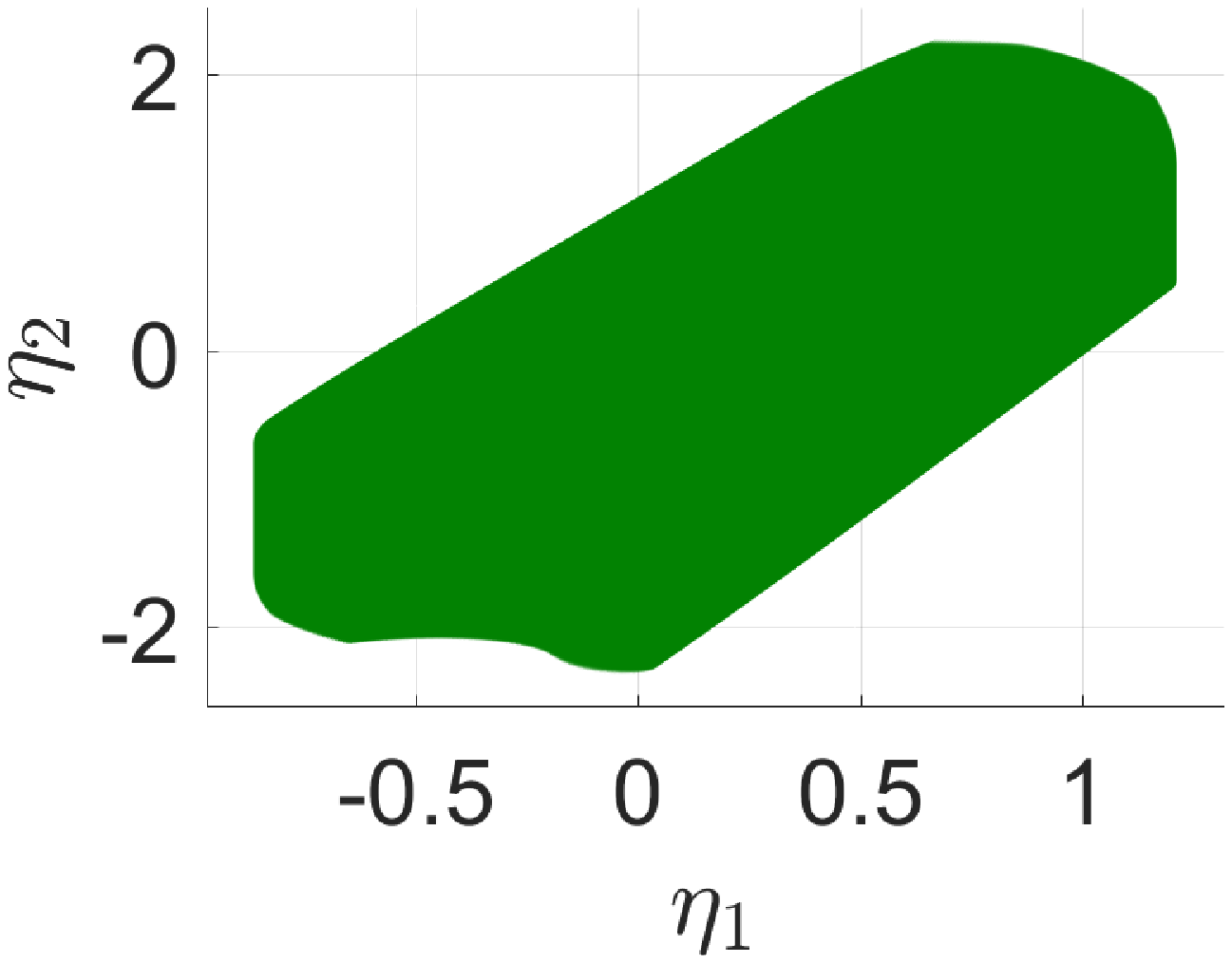}} & \subcaptionbox{ $\mathcal{P}_{z}(O_{\infty})$ \label{fig:measure}}{\includegraphics[width=0.48\linewidth]{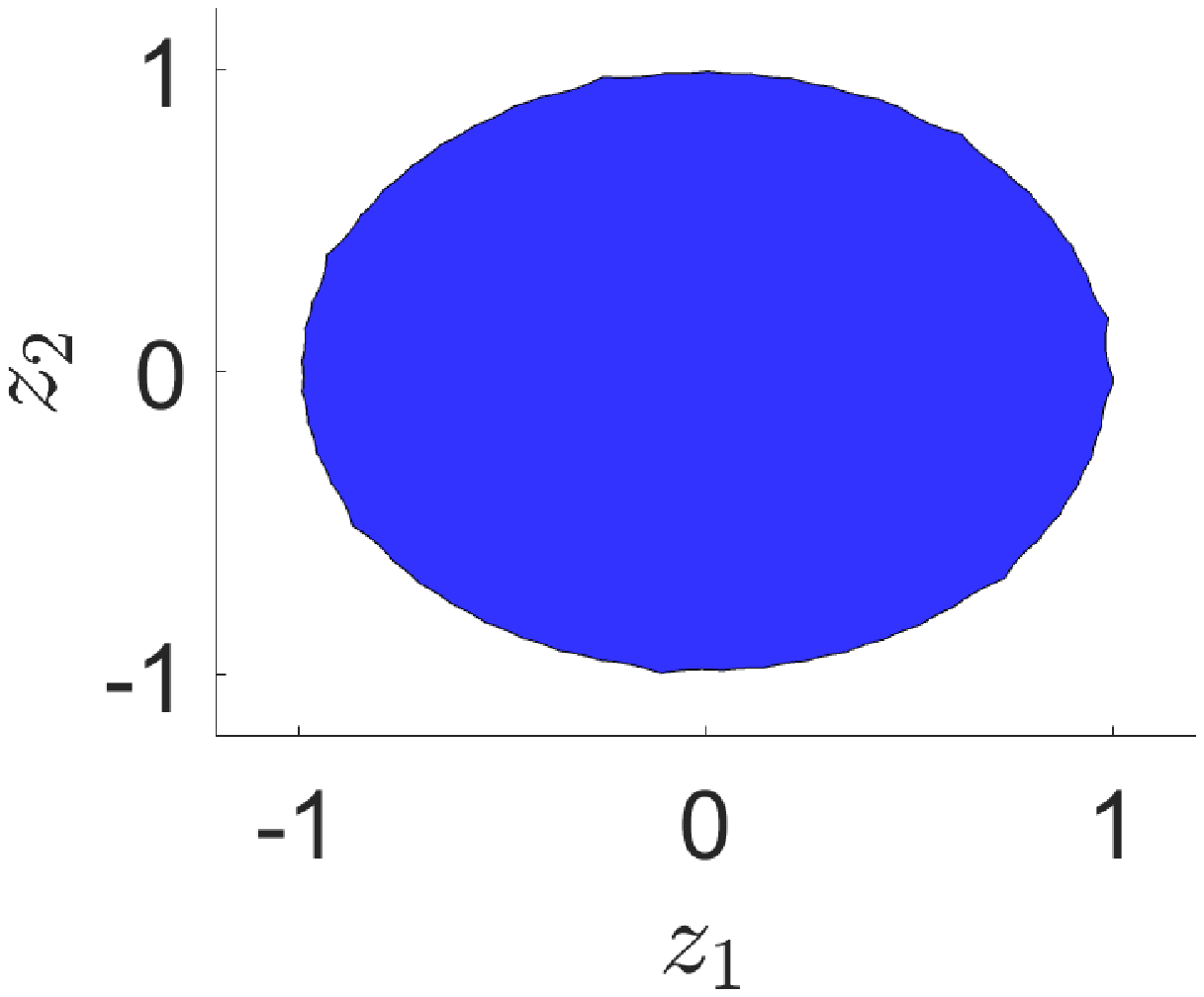}}\\
  \end{tabular}
\caption{Visualization of $O_{\infty}$ of Example 1.}
\label{fig:wiener}
\end{figure}

Now, we consider the problem of computing the \emph{maximal admissible invariant set} of the closed-loop system. The system is subject to the following constraints: $\|\eta\|_{\infty}\le 5, |u|\le 2, \|z\|_{\infty} \le 3, |e|\le 3$. Under the transformation map $T(\eta,z) =\left(
\eta, 
z^{[1]}, 
z^{[2]}, 
z^{[3]}
\right)$, we obtain the lifted linear system
$ \Pi(
\begin{pmatrix}
A_{\eta} + BK  & BL\\
\pmb{0} & A_z^{\pmb{d}}
\end{pmatrix},  \begin{pmatrix}
I_{4} & \pmb{0}
\end{pmatrix}
)
$
where $A_z^{\pmb{d}}$ is defined as in (\ref{eqn:zAd}) with $\pmb{d} = \{1,2,3\}$ and $L = [L_1 ~ L_2 ~ L_3  ]$. Let $O_{\infty}^L$ be the  \emph{maximal admissible invariant set} of the lifted linear system. Then, as shown in Theorem \ref{thm:Omax}, $O_{\infty} = T^{-1}O_{\infty}^L$ is the  \emph{maximal admissible invariant set} of the original system. To visualize this $4$-dimensional set, we plot out its projections on to $\eta$ and $z$, denoted by $\mathcal{P}_{\eta}(O_{\infty})$ and $\mathcal{P}_{z}(O_{\infty})$ respectively, in Figure \ref{fig:wiener}.

\textbf{Example 2} Consider the following double-Zone building thermal model \cite{ART:WH17}:
$
c_i \dot{\mathcal{T}}_i = \frac{\mathcal{T}_{j}-\mathcal{T}_i}{R_{ij}} + \frac{\mathcal{T}_{o}-\mathcal{T}_i}{R^o_{i}} + u_ic_p(\mathcal{T}^s_{i}-\mathcal{T}_i) + q_i, j\not= i, i = 1,2,
$
where $\mathcal{T}_i$ is the temperature of zone $i$, $\mathcal{T}_o$ is the temperature of outside air, $c_i$ is the thermal capacitance of the air in zone $i$, $R_{ij}$ denotes the thermal resistances between zone $i$ and zone $j$, $R^o_i$ denotes the thermal resistance between zone $i$ and the outside environment, $c_p$ is the specific heat capacity of air, $\mathcal{T}^s_{i}$ is the temperature of the supply air delivered to zone $i$, $u_i$ is the flow rate into zone $i$ and $q_i$ is the thermal disturbance from internal loads like occupants and lighting. As the temperature of the supply air is usually constant over short intervals of time, it is assumed to be fixed and known. The outside air temperature here is $\mathcal{T}_o=\SI{38}{\degreeCelsius}$. The thermal disturbance is bounded as: $q_i\in [0.7,0.13], i = 1,2$. Other system parameters are given in the following table.

\begin{table}[H]
\centering
\renewcommand{\arraystretch}{1.5}
\begin{tabular}{|c|c|c|}
\hline
Symbol & Value & Units \\ \hline
$c_1=c_2$ & $1.375\times 10^3$ & $\si{kJ/K}$ \\ \hline
$c_p$ & $1.012$ & $\si{kJ/(kg\cdot K)}$ \\ \hline
$R_{12}=R_{21}$ & $1.5$ & $\si{K/kW}$ \\ \hline
$R^o_1 = R^o_2$ & $3$ & $\si{K/kW}$ \\ \hline
$\mathcal{T}^s_{1}=\mathcal{T}^s_{2}$ & $16$ & $\si{\degreeCelsius}$ \\ \hline
\end{tabular}
\caption{Parameters of the building system} \label{tab:para}
\end{table}

In the simulation, the temperature set-points of zone $1$ and zone $2$ are $\SI{23}{\degreeCelsius}$ and $\SI{24}{\degreeCelsius}$ respectively. Hence,  the steady state is $\mathcal{T}_s = (24,25)$ and the steady control input is $u_s=(0.8140,0.5064)$ for $q_1=q_2=0.1$. The control constraints are: $0 \le u_1 \le 1.5 ,0 \le  u_2 \le 1.5, u_1 + u_2 \le 2$, and the temperature constraints are: $16\le \mathcal{T}_i\le 38, i=1,2$. We discretize the continuous-time system by the zero-order-hold method with the sampling time $\Delta t= \SI{10}{\minute}$ and consider the stabilizing control law $u = K(x-\mathcal{T}_s)+u_s$ with
$
K = [ 0.0633 ~   -0.0756 ; -0.0768 ~  0.0935 ].
$
Let the state be $x=\mathcal{T}-\mathcal{T}_s$ and the disturbance be $d_i=q_i-0.1,i=1,2$. The closed loop system becomes a polynomial system of degree $2$: $x_1^+ =  -0.0279x_1^2+0.0334x_1x_2 + 0.0086x_1 + 0.5246x_2 + 0.4364d_1; x_2^+ = -0.0413x_2^2+0.0339x_1x_2 + 0.5624x_1 + 0.0097x_2 + 0.4364d_2$.
%\begin{align*}
%\left( \begin{array}{c}
%x_1^+ \\
%x_2^+ 
%\end{array}  \right) =& f(x)+Gd=
%\left( \begin{array}{c}
%-0.0279x_1^2+0.0334x_1x_2\\
%-0.0413x_2^2+0.0339x_1x_2
%\end{array}  \right) \\
% +& \left( \begin{array}{cc}
%0.0086 & 0.5246\\
%0.5624 & 0.0097
%\end{array} \right) \left( \begin{array}{c}
%x_1\\
%x_2
%\end{array}  \right) + \left( \begin{array}{c}
%0.4364d_1\\
%0.4364d_2
%\end{array}  \right).
%\end{align*}
From the constraints on the temperature and the input, the state constraint set become $X=\{x\in \mathbb{R}^2: -7\le x_1\le 15,-8\le x_2\le 14, 0 \le Kx+u_s\le 1.5, [1~1] (Kx+u_s)\le 2\}$. The Lipschitz constant $L_f$ of the nominal dynamics $f(x)$ in $X$ can be computed by solving a semidefinite program, see  the appendix for details. We obtain that $L_f=1.5401$. All the assumptions (\textbf{A1})-(\textbf{A4}) are also verified formally in the appendix. 

\begin{figure}[h]
\centering
\includegraphics[width=0.7\linewidth]{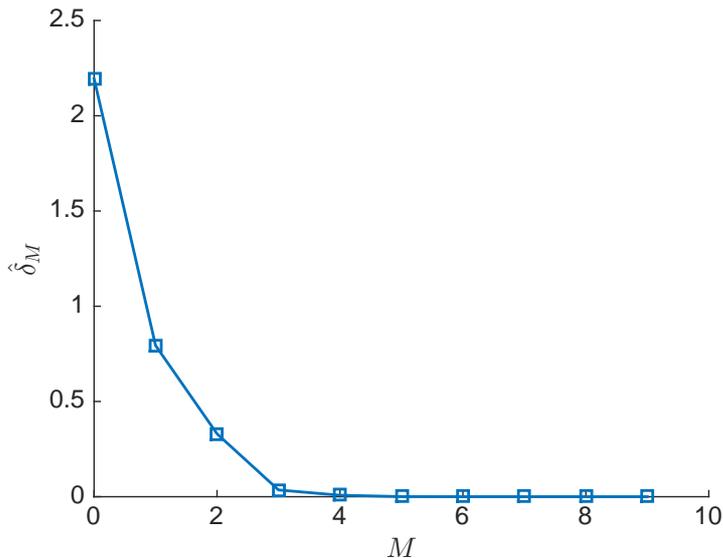}
\caption{Mismatch errors between the lifted system and the original system for different values of $M$ for the double-zone building system.} 
\label{fig:deltaMbuilding}
\end{figure}

\begin{figure}[h]
\centering
\includegraphics[width=0.7\linewidth]{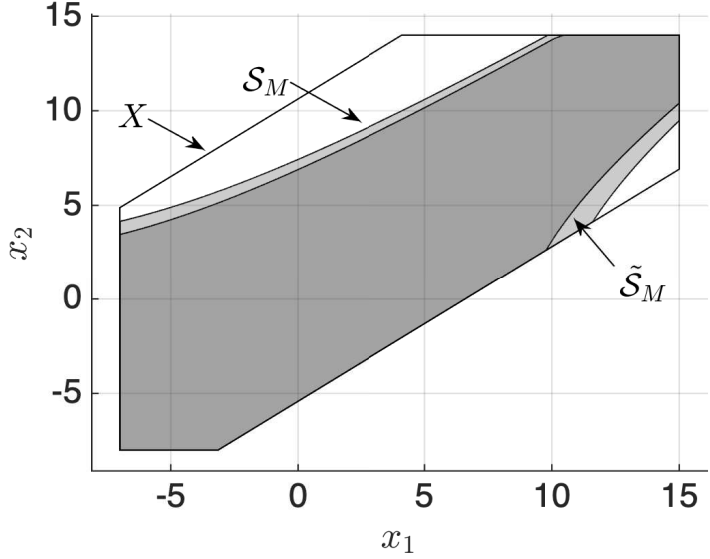}
\caption{Visualization of the sets obtained from Algorithm \ref{algo:datadriven} with $M=5$ for the the double-zone building system.} 
\label{fig:invsetbuilding}
\end{figure}

First, we compute the lifted linear system for the nominal system $x^+=f(x)$ following the procedure in Section \ref{sec:com}. We sample $1.3\times 10^4$ points over $X$ by gridding, solve Problem (\ref{eqn:sumsumgammaM}) and compute $\hat{\delta}_M$ for different values of $M$ as shown in Figure \ref{fig:deltaMbuilding}. As we can see from this Figure, $\hat{\delta}_M$ is already close to $0$ when $M\ge 5$. We set $\delta$ to be $0.01$ in Algorithm \ref{algo:datadriven}. The output is $M=5$ and $\Omega_M = O_{\infty}^{L,\hat{\Delta}_M}(A_{M},C_{M})$ with $\hat{\Delta}_M = B_M \Delta_{\hat{\delta}_M}$. To account for the disturbance, we also compute $\tilde{\Omega}_M = O_{\infty}^{L,\tilde{\Delta}}(A_{M},C_{M})$ with $\tilde{\Delta} = \hat{\Delta}_M + \tilde{L}_M D\|G\|  \mathbb{B}_{12}$, where $\tilde{L}_M = \frac{L_f^{M+1}-1}{L_f-1}$ and $D=0.0131$ is an upper bound on the disturbance of the discretized system. With the transformation $T(x)=\mathcal{F}_5(x)$, we can immediately obtain the sets $\mathcal{S}_M=T^{-1}(\Omega_M)$ and $\tilde{\mathcal{S}}_M=T^{-1}(\tilde{\Omega}_M)$, which are shown in Figure \ref{fig:invsetbuilding}.

\section{Conclusions}\label{sec:con}
We have proposed an immersion-based method for computing the \emph{maximal admissible invariant set} of discrete-time nonlinear systems in a given constraint set. It characterizes the \emph{maximal admissible invariant set} using a lifted linear model. For certain nonlinear systems, exact immersion can be achieved and hence this characterization is also exact. For general cases, the lifted linear system is not exactly equivalent to the nonlinear system and we use the fixed-point iteration technique for invariant sets to compute an \emph{approximate immersion} in a local region of interest. Provided that an upper bound on the mismatch error is available, the proposed characterization can be only considered as an inner approximation of the actual \emph{maximal admissible invariant set}. Nevertheless, we have shown that this inner approximation is an invariant set itself due to a tightening procedure. Finally, the proposed method is demonstrated on two nonlinear examples.

\appendix
\setcounter{secnumdepth}{0}
\section{Appendix: Computational details of Example 2}

\begin{figure}[h]
	\centering
	\includegraphics[width=0.7\linewidth]{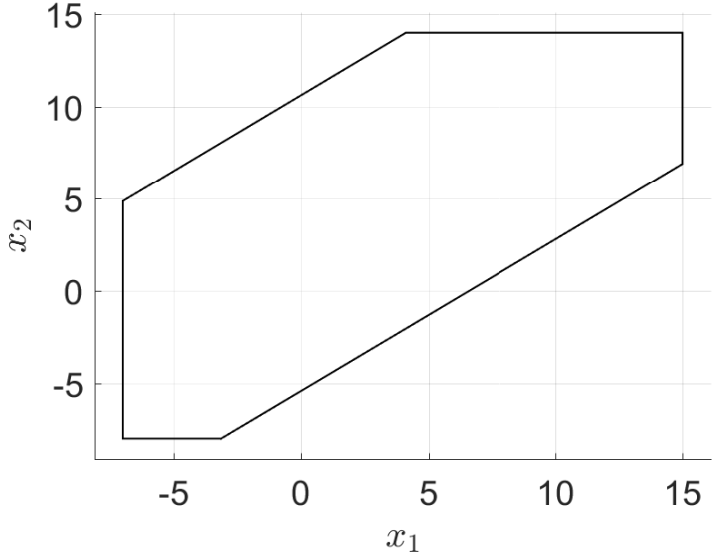}
	\caption{The constraint set $X$ for the double-zone	building system.} 
	\label{fig:X}
\end{figure}

The nominal system is given below:
\begin{align*}
	x_1^+ &=  -0.0279x_1^2+0.0334x_1x_2 + 0.0086x_1 + 0.5246x_2 \\
	x_2^+ &= -0.0413x_2^2+0.0339x_1x_2 + 0.5624x_1 + 0.0097x_2 .
\end{align*}
For convenience, let
\begin{align*}
	A = \begin{pmatrix}
		0.0086 & 0.5246\\
		0.5624 & 0.0097
	\end{pmatrix},
	\bar{A}(x) = \begin{pmatrix}
		-0.0279x_1 + 0.0334x_2 & 0\\
		0 & 0.0339x_1-0.0413x_2
	\end{pmatrix}.
\end{align*}
The system above can then be rewritten as
\begin{align*}
	\begin{pmatrix}
		x_1\\
		x_2
	\end{pmatrix}^+ = f(x) \coloneqq \left(A+\bar{A}(x)\right) \begin{pmatrix}
		x_1\\x_2
	\end{pmatrix}
\end{align*}
The constraint set is 
$X=\{x\in \mathbb{R}^2: -7\le x_1\le 15,-8\le x_2\le 14, 0 \le Kx+u_s\le 1.5, [1~1] (Kx+u_s)\le 2\}$, where $
K = [ 0.0633 ~   -0.0756 ; -0.0768 ~  0.0935 ]
$ and $u_s=[0.8140 ~ 0.5064]$. This set is plotted in Figure \ref{fig:X}.

\textbf{We first show that A1 is satisfied}. This assumption can be easily verified by checking the gradient of $f(x)$:
\begin{align*}
	\nabla f(x) = A + \begin{pmatrix}
		-0.0558x_1+0.0334x_2 & 0.0334x_1\\
		0.0339x_2 & -0.0826x_2+0.0339x_1
	\end{pmatrix}.
\end{align*}
As $X$ is compact, $\|\nabla f(x)\|$ is also bounded for any $x\in X$.  This proves the Lipschitz continuity of the dynamics in $X$. In fact, an explicit bound can be computed by solving  $\max_{x\in X} \|\nabla f(x)\|$. Note that $\nabla f(x)$ is affine in $x$ and $X$ is a (convex) polytope. The maximum is reached at the vertices of $X$, i.e., $\max_{x\in X} \|\nabla f(x)\| =\max_{x\in \mathcal{V}(X)} \|\nabla f(x)\|$, where $\mathcal{V}(X)$ denotes the set of vertices (or extreme points) of $X$. In fact, the maximum of a convex quadratic function over a (convex) polytope is always reached at a vertex. To be self-contained, we provide a proof for this elementary result. 

The problem $\max_{x\in X} \|\nabla f(x)\|$ can be rewritten as a single-variable robust optimization
\begin{align*}
	&\min_{\gamma\ge 0} \gamma\\
	\textrm{s.t. } \quad &\|\nabla f(x)\| \le \gamma, \quad \forall x \in X,
\end{align*}
which is equivalent to 
\begin{align*}
	&\min_{\gamma\ge 0} \gamma\\
	\textrm{s.t. } \quad & \left(\nabla f(x)\right)^\top \nabla f(x)   \le \gamma ^2 I,  \quad  \forall x \in X.
\end{align*}
Using the Schur complement to the problem above yields
\begin{align*}
	&\min_{\gamma \ge 0} \gamma\\
	\textrm{s.t. } \quad &  \begin{pmatrix}
		\gamma^2 I & \nabla f(x)^\top \\
		\nabla f(x) & I
	\end{pmatrix} \succeq 0, \quad \forall x\in X.
\end{align*}
As $\nabla f(x)$ is affine in $x$,  
\begin{align*}
	\begin{pmatrix}
		\gamma^2 I & \nabla f(x)^\top \\
		\nabla f(x) & I
	\end{pmatrix} \succeq 0, \forall x\in \mathcal{V}(X) \Longleftrightarrow
	\begin{pmatrix}
		\gamma^2 I & \nabla f(x)^\top \\
		\nabla f(x) & I
	\end{pmatrix} \succeq 0,  \forall x\in X.
\end{align*}
Thus, the problem above with an infinite number of constraints reduces to 
\begin{align*}
	&\min_{\gamma \ge 0} \gamma\\
	\textrm{s.t. } \quad &  \begin{pmatrix}
		\gamma^2 I & \nabla f(x)^\top \\
		\nabla f(x) & I
	\end{pmatrix} \succeq 0, \quad \forall x\in \mathcal{V}(X).
\end{align*}
Again, using the Schur complement, we arrive at 
\begin{align*}
	&\min_{\gamma\ge 0} \gamma\\
	\textrm{s.t. } \quad & \left(\nabla f(x)\right)^\top \nabla f(x)   \le \gamma ^2 I,  \quad  \forall x \in \mathcal{V}(X),
\end{align*}
which is equivalent to 
\begin{align*}
	&\min_{\gamma\ge 0} \gamma\\
	\textrm{s.t. } \quad &\|\nabla f(x)\| \le \gamma, \quad \forall x \in \mathcal{V}(X).
\end{align*}
With this, we conclude that $\max_{x\in X} \|\nabla f(x)\| =\max_{x\in \mathcal{V}(X)} \|\nabla f(x)\|$.

Finally, we compute the the Lipschitz constant 
\begin{align*}
	L_f = \max_{x\in \mathcal{V}(X)} \|\nabla f(x)\| = 1.5401.
\end{align*}

\textbf{We then show that  A2 is satisfied}. The compactness of $X$ is obvious from Figure \ref{fig:X}. We only need to show that there exists an invariant set with a non-empty interior in $X$. This is done by analyzing reachable sets of the dynamics $x^+ = f(x)$ from the constraint set $X$. Given any $Z\subseteq \mathbb{R}^2$, the one-step forward reachable set from $Z$ is defined as
\begin{align*}
	\mathcal{R}(Z) \coloneqq \{f(x):x\in Z\}.
\end{align*}
For any (convex) polytope $Z\subseteq \mathbb{R}^2$, we also define the following operator:
\begin{align*}
	\overline{\mathcal{R}}(Z) \coloneqq \textrm{conv}\{(A+\bar{A}(v))x: x\in \mathcal{V}(Z), v\in \mathcal{V}(Z)\}.
\end{align*}
Note that $\bar{A}(x)$ is affine in $x$, which means that, $\forall x\in Z$,  $A+\bar{A}(x) \in \textrm{conv}\{A+\bar{A}(v): v\in \mathcal{V}(Z)\}$. Hence, $\forall x\in Z$, $f(x)=(A+\bar{A}(x))x \in \textrm{conv}\{(A+\bar{A}(v))x:  v\in \mathcal{V}(Z)\}$.
It can thus be verified that $\mathcal{R}(Z) \subseteq \overline{\mathcal{R}}(Z)$ for any (convex) polytope $Z$. Hence, $\overline{\mathcal{R}}(Z)$ can be considered as an over-approximation of $\mathcal{R}(Z)$.

\begin{figure}[h]
	\centering
	\includegraphics[width=0.9\linewidth]{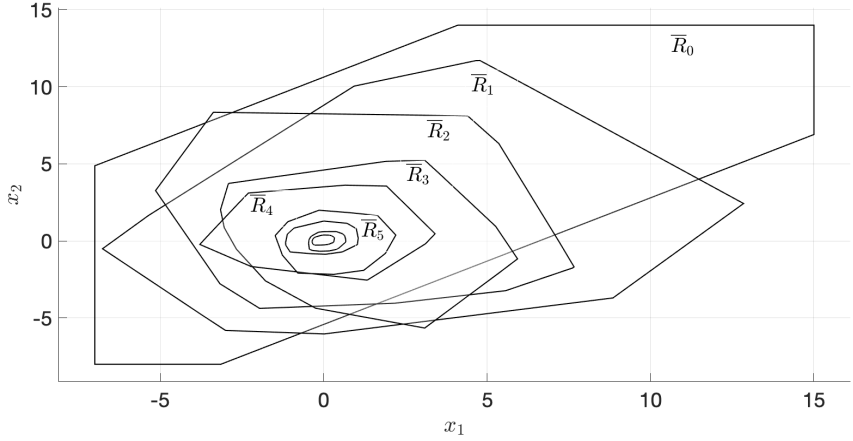}
	\caption{Over-approximations of the reachable sets. } 
	\label{fig:reach}
\end{figure}

Starting from $X$, we then define the following iteration:
\begin{align*}
	R_0 = X,R_{k+1} = \mathcal{R}(R_k), k\ge 0.
\end{align*}
The computation of $\{R_k\}$ is not easy as $f(x)$ is nonlinear. Instead, we use the following iteration:
\begin{align*}
	\overline{R}_0 = X,
	\overline{R}_{k+1} = \overline{\mathcal{R}}\left( \overline{R}_{k} \right), k\ge 0.
\end{align*}
Repeating the same argument above inductively, we conclude that $R_k \subseteq \overline{R}_{k}$ for all $k \ge 0$. The sets $\{\overline{R}_{k}\}$ are plotted in Figure \ref{fig:reach}. It can be seen from this figure that $\overline{R}_{6} \subseteq \overline{R}_{5} \subseteq X$. Note that $\mathcal{R}(\overline{R}_{5}) \subseteq \overline{\mathcal{R}}(\overline{R}_{5}) = \overline{R}_{6} \subseteq \overline{R}_{5}$. Hence, from Figure \ref{fig:reach}, we have the following two observations:

\begin{itemize}
	\item $\overline{R}_{5}$ is an invariant set contained in $X$. 
	\item $f^t(x)\in \overline{R}_{5} \subseteq X$ for any $t\ge 5$ and any $x\in X$.
\end{itemize}

Therefore, A2 is satisfied.

\textbf{We now show that A3 is also satisfied with $\mathcal{A}=\{0\}$}. More precisely, we want to show that there exists a class $\mathcal{K}\mathcal{L}$ function $\beta$ such that  $
\|f^t(x)\| \le \beta(\|x\|,t), \forall t\in \mathbb{Z}^+, \forall x\in X$. First, we compute an upper bound of the norm of $A+\bar{A}(x)$  for all $x\in \overline{R}_{5}$ by solving $\max_{x\in \overline{R}_{5}} \|A+\bar{A}(x)\|$. As $\bar{A}(x)$ is affine in $x$ and  $\overline{R}_{5}$ is a (convex) polytope, following the arguments above, the maximum is $\max_{x\in \mathcal{V}(\overline{R}_{5})} \|A+\bar{A}(x)\|$, denoted as $\rho(\overline{R}_{5})$.  We obtain that $\rho(\overline{R}_{5}) = 0.5822$, which means that  $\|A+\bar{A}(x)\| \le 0.5822$ for any $x\in \overline{R}_{5}$. By the invariance of $ \overline{R}_{5}$, we know that $\|f^{t}(x)\|\le 0.5822^{t-5} \|f^5(x)\|$ for any $x\in X$ and any $t \ge 5$. Similarly, we also compute $\rho(\overline{R}_{i})$ for $i=0,1,\cdots, 4$. Finally, we obtain that, for any $x\in X$ and $t\ge 5$,
\begin{align*}
	\|f^{t}(x)\| &\le 0.5822^{t-5} \|f^5(x)\| \\
	& \le 0.5822^{t-5} \rho(\overline{R}_{4}) \|f^4(x)\| \\
	& \le 0.5822^{t-5} \rho(\overline{R}_{4}) \rho(\overline{R}_{3}) \|f^3(x)\| \\
	& \le 0.5822^{t-5} \rho(\overline{R}_{4}) \rho(\overline{R}_{3}) \rho(\overline{R}_{2})\|f^2(x)\| \\
	& \le 0.5822^{t-5} \rho(\overline{R}_{4}) \rho(\overline{R}_{3}) \rho(\overline{R}_{2}) \rho(\overline{R}_{1})\|f(x)\| \\
	& \le 0.5822^{t-5} \rho(\overline{R}_{4}) \rho(\overline{R}_{3}) \rho(\overline{R}_{2}) \rho(\overline{R}_{1}) \rho(\overline{R}_{0})\|x\|. 
\end{align*}
Thus, there exists a constant $c$ such that $	\|f^{t}(x)\|  \le c0.5822^t \|x\|$ for any $x\in X$ and $t \ge 0$.

\textbf{Finally, we show that A4 is satisfied}. We compute the Lipschitz constant of $f(x)$ in $\rho(\overline{R}_{5})$ by solving $\max_{x\in \mathcal{V}(\overline{R}_{5})} \|\nabla f(x)\|$. 
Let the solution be denoted as $\mathcal{L}(\overline{R}_{5})$. We obtain that $\mathcal{L}(\overline{R}_{5}) = 0.6546$.  We can also compute $\mathcal{L}(\overline{R}_{i})$ in the same way for any $i=0,1,\cdots, 4$. Thus, for any $x,y\in X$ and $t\ge 6$,
\begin{align*}
	\|f^t(x)-f^t(y)\| &\le 0.6546^{t-5} \|f^5(x)-f^5(y)\|\\
	&\le 0.6546^{t-5} \mathcal{L}(\overline{R}_{4})  \|f^4(x)-f^4(y)\| \\
	& ~~ \vdots\\
	&\le  0.6546^{t-5} \mathcal{L}(\overline{R}_{4})\mathcal{L}(\overline{R}_{3})\mathcal{L}(\overline{R}_{2})\mathcal{L}(\overline{R}_{1}) \mathcal{L}(\overline{R}_{0})  \|x-y\|
\end{align*}
We then conclude that there exists a constant $c$ such that $\|f^t(x)-f^t(y)\| \le c\|x-y\|$ for any $x,y\in X$ and any $t \ge 0$.

\bibliographystyle{unsrt}
\bibliography{Reference}

\end{document}